\documentclass[aps,prx,reprint,superscriptaddress,nofootinbib,longbibliography]{revtex4-2}
\usepackage[T1]{fontenc}
\usepackage{amsmath,amssymb,mathtools,bm,amsthm}
\usepackage{graphicx}
\usepackage[version=4]{mhchem}
\usepackage{hyperref}
\usepackage{booktabs}
\usepackage{tikz}
\usepackage{enumitem}
\usepackage{multirow}
\hypersetup{hidelinks}
\usepackage{multirow}
\usepackage{bbold}
\newcommand{\epsM}{\varepsilon^2 M}
\newcommand{\Var}{\operatorname{Var}}
\newcommand{\Cov}{\operatorname{Cov}}
\newcommand{\calG}{\mathcal{G}}
\newcommand{\calC}{\mathcal{C}}

\newcommand{\Eh}{E_{\rm h}}

\begin{document}

\title{Binary Optimization of Measurement Groupings for Quantum Energy Estimation}
\author{Isaac L. Huidobro-Meezs}
\email{huidobri@mcmaster.ca}
\affiliation{Department of Chemistry \& Chemical Biology, McMaster University, Hamilton, ON Canada}

\author{Rodrigo A. Vargas-Hern\'andez}
\email{vargashr@mcmaster.ca}
\affiliation{Department of Chemistry \& Chemical Biology, McMaster University, Hamilton, ON Canada}
\affiliation{School of Computational Science \& Engineering, McMaster University, Hamilton, ON Canada}
\affiliation{Brockhouse Institute for Materials Research, McMaster University, Hamilton, ON Canada}

\date{\today}

\begin{abstract}
Repeated measurements can dominate the resources required for quantum energy estimation, making the choice of which Pauli observables to measure together a central optimization problem for variational quantum algorithms. We formulate fully commuting measurement grouping as a classical binary optimization problem based on clique selection and construct non-overlapping groups using mixed-integer linear programming (MILP). For a benchmark of molecular Hamiltonians, groupings optimized with approximate covariances reduce the non-overlapping measurement requirement $\varepsilon^2M$ by $51.8\%$ on average relative to sorted insertion (SI). Using the MILP groups to initialize iterative coefficient splitting (ICS), denoted MILP-ICS, yields an average $24.3\%$ reduction relative to ICS initialized from SI (SI-ICS). The optimized groups also transfer across nearby molecular geometries while preserving substantial measurement savings. We further introduce O-clique, which directly optimizes overlapping commuting supports through candidate-clique selection and coefficient profiles. Although O-clique provides only modest additional reductions beyond MILP-ICS, it reduces the measurement requirement by $27.3\%$ on average relative to SI-ICS with 100 iterations, despite using only five final coefficient refinement iterations, indicating improved support quality. Finally, we extend the comparison to Fermi--Hubbard, Kitaev--Heisenberg--$\Gamma$, and XYZ lattice Hamiltonians, where MILP-based groupings substantially outperform the corresponding SI-based strategies. Together, these results show that variance-informed optimization of measurement-group structure can substantially reduce sampling costs across molecular and lattice Hamiltonians, with optimized non-overlapping groups providing strong, transferable initializations and direct overlapping optimization offering further gains.
\end{abstract}

\maketitle
\raggedbottom

\section{Introduction}
\label{sec:intro}
Preparing a quantum state in a quantum device is not sufficient to determine its energy. The energy must be inferred from repeated measurements, and the repetitions needed to reach a useful precision can dominate the resources required for a calculation. This sampling cost is a major obstacle to practical quantum chemistry and many-body physics applications on quantum devices and remains an essential consideration when assessing what quantum devices can accomplish~\cite{MeasurementsRoadblock,Preskill2018NISQ,Patel2025Review}. Improving measurement strategies can therefore make an algorithm more useful without changing the state it prepares: it obtains the same physical information with fewer executions of the quantum circuit.

The challenge extends across algorithms that extract information from expectation values. Variational approaches repeatedly estimate energies during classical optimization~\cite{Peruzzo2014VQE,McClean2016Theory,VQARev}; related applications include adaptive molecular ans\"atze~\cite{ADAPT-VQE}, excited-state calculations~\cite{VarExcited,QSE,KrylovSubspace}, and contracted-eigenvalue methods~\cite{ContractedSchrodinger}. Measurements are also central to variational studies of gauge theories and frustrated quantum systems~\cite{Zhang2023simulatinggauge,PhysRevB.102.075104}, as well as quantum machine learning and regression~\cite{QuantumBoltzmannML,QuantumCircuitLearning,MeasurementQuantumNN,DeepQuantumNNs,RegressionQuantum}. In each setting, the cost of one expectation-value estimate is compounded when the estimate must be repeated throughout a larger calculation.

For electronic structure, the qubit Hamiltonian contains many Pauli terms whose contributions must be combined to recover the energy~\cite{BKLove}. Measuring every term separately is unnecessary: compatible terms can be measured together, so one circuit can yield several expectation values. This observation motivates graph-based partitions and efficient greedy constructions such as sorted insertion (SI)~\cite{Verteletskyi2020MinimumClique,Crawford2021SortedInsertion}, as well as learned approaches to grouping~\cite{HuidobroMeezs2024GFlowNets,huidobromeezs2026discreteflowbasedgenerativemodels}. Fewer groups, however, do not necessarily mean fewer measurements. The statistical fluctuations within each group, including correlations between its terms, determine how many repetitions are needed for the desired energy precision~\cite{OrigErrorExpression,Romero_2019,Crawford2021SortedInsertion}.

A further reduction is possible when a Pauli term can contribute to more than one measurement group. Overlapping strategies distribute the coefficients of Pauli operators from the qubit Hamiltonian among compatible groups while preserving the original Hamiltonian~\cite{Yen2023Deterministic,Wu2023OverlappedGrouping,rowland2026overlappedgroupingsRepacking}. This additional freedom, together with related fragment-reconstruction strategies, can reduce the variance beyond that of a non-overlapping partition~\cite{Choi2022Ghost,Choi2023Fluid}. Iterative coefficient splitting (ICS) \cite{Yen2023Deterministic} and related overlapping methods \cite{Choi2022Ghost,Choi2023Fluid} use a non-overlapping initialization to form overlapping groups, then adjust the coefficients and measurement allocations. The initial groups determine which combinations of compatible terms are available to the subsequent optimization, affecting the downstream performance of these methods~\cite{VarSI}.

The information available during measurement design is another practical consideration. State-independent heuristics are attractive due to their simplicity \cite{Crawford2021SortedInsertion,rowland2026overlappedgroupingsRepacking}; however, their measurement counts can be improved. Variance-informed strategies use estimates of the state-dependent correlations, which can be obtained from an approximate state or refined as measurement data becomes available~\cite{Yen2023Deterministic,Shlosberg2023adaptiveestimation,VarSI}.

This dependence raises two connected questions: i) how should non-overlapping groups be chosen when they will ultimately initialize an overlapping estimator, and ii) is it more effective to improve that initialization or to search directly for overlapping groups? An optimal non-overlapping measurement score does not, by itself, determine the quality of the final coefficient-split estimator. Likewise, coefficients that are nearly optimal for one overlapping grouping do not establish that the grouping itself is a good choice. Resolving these distinctions requires comparing the final measurement requirements.

Here, we cast measurement-group design as a classical binary optimization problem and distinguish three levels of optimization: 
\begin{enumerate}[label=\Roman*., leftmargin=*]
    \item Non-overlapping groups construction.
    \item Overlapping coefficient optimization from variance-informed initializations.
    \item Direct search over overlapping measurement supports.
\end{enumerate}
We first formulate non-overlapping measurement design as a candidate-clique selection problem and solve it using mixed-integer linear programming (MILP) to construct variance-informed, fully commuting groups. We evaluate these groups by both measurement cost and performance as initializations for ICS. We then introduce O-clique, which selects overlapping candidate cliques jointly with coefficient heuristics to determine whether direct support optimization improves on the MILP-ICS solution. For molecular Hamiltonians, we test the robustness of these strategies using approximate CISD covariances during optimization and by transferring optimized groupings across nearby geometries. We further extend the comparison to Fermi--Hubbard, Kitaev--Heisenberg--$\Gamma$, and XYZ lattice Hamiltonians, whose interaction and covariance structures differ qualitatively from those of molecular electronic-structure problems. These benchmarks separate gains from measurement-support selection from those from coefficient refinement, and quantify when the additional classical cost of optimizing the overlapping support is warranted.

\section{Methods}
\label{sec:methods}
Consider a Hamiltonian on $N_q$ qubits,
\begin{equation}
H=c_0 \mathbb{1}+\sum_{i=1}^{N_P}c_iP_i,
\label{eq:hamiltonian}
\end{equation}
where the $c_i$ are real coefficients, $\mathbb{1}$ is the identity, and the $P_i$ are distinct non-identity Pauli words. A measurement grouping $\calG$ specifies the decomposition,
\begin{equation}
H=c_0 \mathbb{1}+\sum_{\alpha=1}^{N_f}H_\alpha,
\qquad H_\alpha=\sum_{i\in G_\alpha}\omega_i^{(\alpha)}P_i,
\label{eq:partition}
\end{equation}
with coefficient reconstruction,
\begin{equation}
c_i=\sum_{\alpha\in I_i}\omega_i^{(\alpha)}.
\label{eq:sumCoeff}
\end{equation}
Here, $I_i$ is the set of groups containing $P_i$, and $N_f$ denotes the available group labels, including empty labels during a search. We define $N_G(X)$ as the number of groups. A non-overlapping grouping corresponds to $|I_i|=1$ and $\omega_i^{(\alpha)}=c_i$ if the Pauli operator $i$ appears on group $\alpha$. 

Starting from the qubit Hamiltonian of Eq. \ref{eq:hamiltonian}, we define a commutativity graph $G_{\mathrm{comm}}=(V,E_{\mathrm{comm}})$ with one vertex $v_k\in V$ for each Pauli word $\hat{P}_k$ and the edge set of $G_{\mathrm{comm}}$: $(i,j)\in E$, with $i<j$, when $[P_i,P_j]=0$; see Fig.~\ref{fig:main_graph}. Each group corresponds to a clique in the commutativity graph. We define a binary variable denoting the membership of each Pauli operator to a group as,
\begin{equation}
x_{i,\alpha}=\begin{cases}
1,&P_i\in G_\alpha,\\
0,&P_i\notin G_\alpha,
\end{cases}
\qquad X=(x_{i,\alpha}).
\label{eq:binaryVariables}
\end{equation}
Thus $I_i=\{\alpha:x_{i,\alpha}=1\}$ and the multiplicity of a Pauli operator is $m_i=|I_i|$. The common feasibility conditions are,
\begin{align}
1\leq \sum_\alpha x_{i,\alpha}&\leq m_i^{\max},\label{eq:coverage}\\
x_{i,\alpha}+x_{j,\alpha}&\leq1\quad(i<j,\ (i,j)\notin E),\label{eq:compatibility}\\
x_{i,\alpha}=0&\ \Longrightarrow\ \omega_i^{(\alpha)}=0.\label{eq:membership_coefficients}
\end{align}
The bound $m_i^{\max}=|I_i|_{\max}$ limits the number of occurrences of a term. Non-overlapping groupings correspond to $m_i^{\max}=1$. This problem admits an equivalent constraint-energy Ising representation \cite{IsingNPProblems}, and we give its forms in Appendix~\ref{app:ising}. We do not substitute them for these constraints in our MILP calculations as the number of variables becomes unfavorable due to the quartic scaling of Pauli words with the number of qubits. However, they provide, in combination with a quadratic reward proxy, a potential test case for Quantum Approximate Optimization Algorithm (QAOA) based approaches \cite{QAOA,SCOOP,Ulrike026qubitefficientgateefficientencodingsgraph,UlrikeCombOptTowardsQADv,PenaltyFreeCO_QAOA,QAOAbeatsSDP}.

For a fixed state $\psi$, the covariance of compatible, unit-coefficient Pauli words is
\begin{equation}
C_{ij}=\Cov_\psi(P_i,P_j)
=\langle P_iP_j\rangle_\psi-\langle P_i\rangle_\psi\langle P_j\rangle_\psi.
\label{eq:cov-def}
\end{equation}
Only diagonal entries and compatible-pair covariances are needed to evaluate feasible fragments.

\subsection{Measurement objective for non-overlapping groupings}
\label{sec:nonoverlap}
For a non-overlapping grouping, $\omega_i^{(\alpha)}=c_i x_{i,\alpha}$, and the fragment variance is,
\begin{equation}
V_{G_\alpha}=\Var_\psi\!\left(\sum_{i\in G_\alpha}c_iP_i\right)
=\sum_{i,j\in G_\alpha}c_ic_jC_{ij}.
\label{eq:group-var}
\end{equation}
If group $\alpha$ receives a fraction $p_\alpha$ of $M$ independent measurements, the energy-estimator variance is $M^{-1}\sum_\alpha V_{G_\alpha}/p_\alpha$. Minimization over $p_\alpha\geq0$, with $\sum_\alpha p_\alpha=1$, gives the number of measurements required to reach a target standard error $\varepsilon$ after optimizing the allocation for that grouping~\cite{OrigErrorExpression,Romero_2019,Yen2023Deterministic}
\begin{align}
S(\calG)&=\sum_{G_\alpha\in\calG}\sqrt{V_{G_\alpha}},\label{eq:score}\\
\epsM(\calG)&=S(\calG)^2,
\qquad p_\alpha=\frac{\sqrt{V_{G_\alpha}}}{S(\calG)}.
\label{eq:intro-objective}
\end{align}
Throughout the paper, we use $\epsM$ as the main metric for comparing the methods.

\begin{figure*}[t!]
    \centering
    \includegraphics[width=0.99\linewidth]{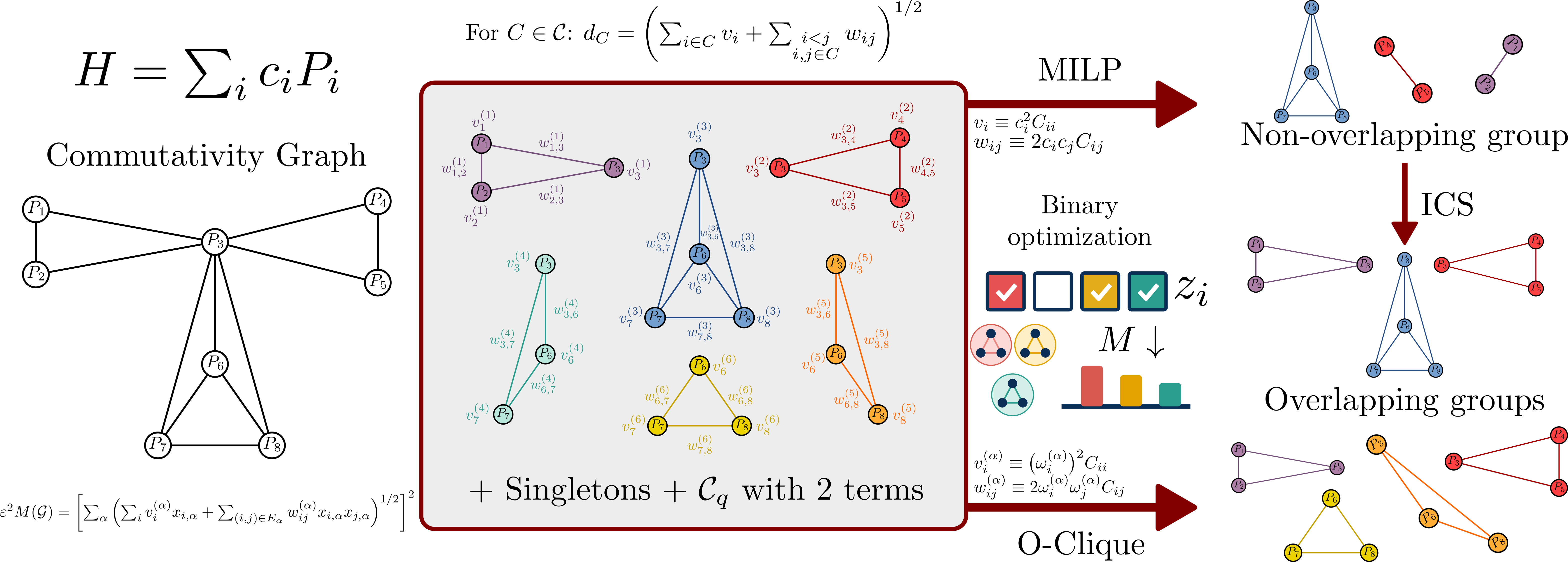}
    \caption{Pipeline for the main algorithms proposed here. Given a Hamiltonian and its commutativity graph, we produce a set of candidate cliques $\mathcal{C}$ including singletons. For each clique, we calculate its contribution to the measurement cost $d_C$ using an approximate covariance dictionary. The MILP variant uses the node and edge weight definitions from Eq. \ref{eq:Node_Edge_weights} and produces a non-overlapping grouping. This non-overlapping result can be employed with ICS to produce an overlapping grouping. The O-Clique variant replaces the node and edge weights with their overlapping definition in Eq. \ref{eq:Node_Edge_weights_overlapping} and employs a heuristic to initialize the weights assigned to each group, producing an overlapping measurement setting. Once it produces an overlapping grouping, O-Clique refines the weights using the same optimization as ICS.}
    \label{fig:main_graph}
\end{figure*}

To expose the graph structure of the objective, we define the \emph{node-like} features and \emph{edge-like} features
\begin{equation}
v_i\equiv c_i^2C_{ii},
\qquad w_{ij}\equiv 2c_ic_jC_{ij}.
\label{eq:Node_Edge_weights}
\end{equation}
The node feature ($v_i$) is the variance of the isolated Hamiltonian term, whereas the edge feature ($w_{ij}$) is its pairwise covariance contribution; see Fig.~\ref{fig:main_graph}. The associated commutativity graph of group $\alpha$, $G_\alpha$, has $N_\alpha$ vertices and $E_\alpha=\{(i,j)\in E: i,j\in G_\alpha\}$ internal edges; thus, its contribution to the measurement cost is given by
\begin{equation}
S(\calG)=\sum_{G_\alpha\in\calG}
\left(\sum_{i\in N_\alpha}v_i+\sum_{(i,j)\in E_\alpha}w_{ij}\right)^{1/2}.
\label{eq:node_edge_score}
\end{equation}
Equivalently, on the feasible binary domain, using $x_{i,\alpha}^2=x_{i,\alpha}$,
\begin{equation}
\small
\epsM(X)= \Biggl[\sum_\alpha\Biggl(\sum_i v_ix_{i,\alpha}+\sum_{(i,j)\in E}w_{ij}x_{i,\alpha}x_{j,\alpha} \Biggr)^{1/2}\Biggr]^2.
\label{eq:Non-overlappingObjective}
\end{equation}
The binary products select the internal edges, so summing over $E$ is equivalent to summing over each $E_\alpha$. Although every fragment variance is quadratic in $X$, its square root makes the complete measurement objective generally non-quadratic.\\

A candidate-clique representation avoids this nonlinear dependence during non-overlapping selection and reduces the number of binary variables during optimization by capping the number of candidates considered. Let $\calC$ be a finite pool of compatible cliques, and let $z_C$ indicate whether $C\in\calC$ is selected. 
We assign each candidate its exact standard deviation,
\begin{equation}
d_C=\left(\sum_{i\in C}v_i+\sum_{\substack{i<j\\i,j\in C}}w_{ij}\right)^{1/2}.
\label{eq:binmopt-clique-cost}
\end{equation}
Once a candidate clique $C$ is specified, we can compute $d_C$ before optimization using approximate covariances, making it a constant in the subsequent selection problem, where the optimizer decides whether to include it. 
The restricted optimization is
\begin{equation}
\begin{aligned}
\min_{\{z_C\}}\quad&\sum_{C\in\calC}d_Cz_C,\\
\text{subject to}\quad&\sum_{C\ni i}z_C=1\quad(i=1,\ldots,N_P),\\
&\sum_{C\in\calC}z_C\leq N_f,\qquad z_C\in\{0,1\}.
\end{aligned}
\label{eq:binmopt-clique-master}
\end{equation}
For every feasible selection, the objective equals $S(\calG)$ exactly subject to the covariance dictionary employed. Since $S\geq0$, it has the same minimizers as $\epsM$ on that pool. The number of binary variables is $|\calC|$, instead of $N_PN_f$. The sum $\sum_{C\ni i}$ runs over candidate cliques in $\calC$ containing term $i$, so the coverage equalities ensure that every Pauli word belongs to exactly one selected clique, ensuring that term $i$ occurs in exactly one selected clique. Candidates in the pool may overlap with one another, but two overlapping candidates cannot both be selected. The second constraint limits the number of selected groups to $N_f$. \\

We construct the candidate pool from coloring-based and ordered greedy partitions, supplemented by compatible clique growth. We retain a complete feasible partition within the group bound. For coloring algorithms, we use the complement of the commutativity graph; a color class of the complement graph forms a clique of the commutativity graph. We further refine the selected and seed partitions by single-term relocations, accepted only when they lower the complete $\epsM$. We label the combined construction as MILP. The linear integer subproblems use SciPy's interface to HiGHS~\cite{Huangfu2018MILP}. When certified by the solver, optimality applies to $\epsM$ among feasible non-overlapping partitions from the candidate pool, without guaranteeing global optimality. Appendix~\ref{app:numerical} details candidate construction and incumbent selection. We also considered a hybrid procedure that adds large-neighborhood search (LNS) refinement. Additionally, overlapping methods like ICS often require a non-overlapping initialization. Initializations informed by measurement costs can improve the performance of these overlapping methods \cite{VarSI,huidobromeezs2026discreteflowbasedgenerativemodels}. The resulting MILP partition provides a variance-informed initialization for ICS and related overlapping techniques. Before coefficient optimization, we consider Pauli words in decreasing order of coefficient magnitude and add them to a group whenever they commute with every current member of the destination group. Newly added occurrences initially have zero coefficients, so the expansion preserves the original fragments. ICS then redistributes coefficients among the expanded groups while preserving the Hamiltonian and updates the measurement allocations. We denote this procedure by MILP-ICS.

\subsection{Overlapping clique selection}
\label{sec:overlap}

Overlapping groupings allow a Pauli word to appear in several compatible measurement groups. In the graph representation, this corresponds to allowing the same vertex to belong to multiple cliques, as illustrated in Fig.~\ref{fig:main_graph}. The binary memberships specify which Pauli words belong to each group, while the split coefficients $\omega_i^{(\alpha)}$ determine the contribution of each occurrence to the corresponding measurement fragment. The reconstruction condition,
\begin{equation}
\sum_{\alpha\in I_i}\omega_i^{(\alpha)}=c_i
\label{eq:sumCoeff_overlap}
\end{equation}
ensures that the overlapping fragments reproduce the original Hamiltonian.

For an overlapping grouping, the node-like and edge-like features become group-dependent,
\begin{equation}
v_i^{(\alpha)} \equiv \bigl(\omega_i^{(\alpha)}\bigr)^2 C_{ii},\quad
w_{ij}^{(\alpha)} \equiv 2\omega_i^{(\alpha)}\omega_j^{(\alpha)}C_{ij}.
\label{eq:Node_Edge_weights_overlapping}
\end{equation}
The variance of group $G_\alpha$ is then,
\begin{equation}
V_{G_\alpha}
=
\sum_i v_i^{(\alpha)}x_{i,\alpha}
+
\sum_{(i,j)\in E}
 w_{ij}^{(\alpha)}x_{i,\alpha}x_{j,\alpha},
\label{eq:overlap_group_variance}
\end{equation}
and optimal shot allocation gives
\begin{equation}
\epsM(X,\omega)
=
\left(
\sum_\alpha \sqrt{V_{G_\alpha}}
\right)^2.
\label{eq:OverlappingObjective}
\end{equation}
Specifying a set of overlapping cliques does not determine the overall measurement cost, as the coefficients can be adjusted. The overlapping problem therefore requires both a set of compatible groups and a coefficient decomposition satisfying Eq.~\eqref{eq:sumCoeff_overlap}.

We construct these groupings through overlapping clique selection, denoted O-clique. Rather than optimizing every entry of the membership matrix directly, O-clique selects from a finite pool of candidate cliques, each containing mutually commuting Pauli words,
\begin{equation}
\mathcal{C}
=
\left\{\mathcal{C}_q\right\}_{q=1}^{N_C}.
\label{eq:oclique_candidate_set}
\end{equation}
O-clique aims to produce a complete cover, meaning a collection of compatible cliques whose union contains every Hamiltonian term, where its cliques may overlap. For each candidate clique $\mathcal{C}_q$, a binary variable $y_q\in\{0,1\}$ indicates whether that clique is selected. If the non-overlapping MILP initialization contains $K$ groups, the O-clique selection satisfies
\begin{equation}
\sum_{q=1}^{N_C} y_q \leq K.
\label{eq:oclique_group_cap}
\end{equation}
This permits O-clique to change the group composition and coefficient sharing while bounding the number of selected measurement groups by the MILP initialization. The bound does not require using all $K$ groups, so solutions with fewer groups can be considered.

The candidate pool contains the cliques from the MILP and sorted-insertion partitions, together with their compatible expansions. We generate additional complete covers from alternative clique-growth orderings, graph-coloring and first-fit constructions, randomized feasible covers, and exchange moves that preserve commutativity and complete Hamiltonian coverage. These procedures produce both complete candidate covers and additional compatible cliques that the O-clique optimization can recombine. SI-ICS and MILP-ICS serve as comparisons for the proposed O-clique method. We also retain the MILP-ICS decomposition as a feasible reference throughout the O-clique search. Appendix~\ref{app:numerical} provides further details on candidate generation.

To represent possible coefficient decompositions within the candidate cliques, each clique $\mathcal{C}_q$ is associated with a finite set of coefficient profiles indexed by
\begin{equation}
\mathcal{P}_q = \left\{1,\ldots,N_q^{\mathrm{prof}}\right\}
\end{equation}
A profile $r\in\mathcal{P}_q$ defined as $\mathbf{a}^{(q,r)}\in\mathbb{R}^{N_P}$, specifies one coefficient vector for that clique and has zero entries outside $\mathcal{C}_q$. One profile therefore defines a fragment $A_{qr}=\sum_{i\in\calC_q}\mathbf{a}^{(q,r)}P_i$; a collection of profiles must satisfy coefficient reconstruction to define a decomposition of the full Hamiltonian. Let $\mathbf{a}_q^{(r)}=(a_i^{(q,r)})_{i\in\mathcal{C}_q}$ denote its restriction to the clique, and let $C_q=(C_{ij})_{i,j\in\mathcal{C}_q}$ be the corresponding covariance submatrix. The profile standard deviation is 
\begin{equation}
\sigma_{qr}=\sqrt{\left(\mathbf{a}^{(q,r)}\right)^TC\mathbf{a}^{(q,r)}}=\sqrt{\left(\mathbf{a}_q^{(r)}\right)^TC_q\mathbf{a}_q^{(r)}}.
\label{eq:oclique_profile_std}
\end{equation}

Because every profile is zero outside a compatible clique, Eq.~\eqref{eq:oclique_profile_std} is equivalent to evaluating the corresponding clique covariance submatrix.

The profile set includes the MILP-ICS decomposition, profiles constructed from complete candidate covers, and the raw Hamiltonian-coefficient profile on every candidate clique, with $a_i=c_i$ for $i\in\mathcal{C}_q$ and zero otherwise. 
For each complete cover, we first transfer the MILP-ICS coefficients to its cliques. If $G_\alpha^{\mathrm{ref}}$ is a MILP-ICS support, its similarity to destination clique $\mathcal{C}_q$ is $|G_\alpha^{\mathrm{ref}}\cap\mathcal{C}_q|/|G_\alpha^{\mathrm{ref}}\cup\mathcal{C}_q|$. For each occurrence of $P_i$, we transfer its coefficient to the destination clique containing $i$ that maximizes this similarity, and we add signed contributions assigned to the same occurrence. Transport preserves reconstruction because every source contribution is assigned once. The transported coefficients initialize one of three coefficient-assignment heuristics: relative standard deviation (RSD), relative variance (RV), or score softmax (SM). Each reported variant uses one guide for constructing its profiles.

For the heuristic updates, consider a fixed grouping at iteration $t$. Removing the current contribution of $P_i$ from group $G_\alpha$ defines the background fragment and its variance,
\begin{equation}
H_{\alpha\setminus i}^{(t)}
=\sum_{\substack{j\in G_\alpha\\j\neq i}}\omega_{j,t}^{(\alpha)}P_j,,
\qquad
B_{i,\alpha}^{(t)}
=
\Var_\psi\!\left(H_{\alpha\setminus i}^{(t)}\right).
\label{eq:background_fragment}
\end{equation}
The covariance between the unit-coefficient Pauli word and this background is
\begin{equation}
b_{i,\alpha}^{(t)}=\Cov_\psi\!\left(P_i,H_{\alpha\setminus i}^{(t)}\right)=\sum_{\substack{j\in G_\alpha\\j\neq i}}C_{ij}\omega_{j,t}^{(\alpha)}.
\label{eq:background_covariance}
\end{equation}
At iteration $t$, a nonnegative guide weight $r_{i,\alpha}^{(t)}$ is assigned to each $\alpha\in I_i$. Normalizing these weights gives the target coefficient
\begin{equation}
\widetilde{\omega}_{i,t}^{(\alpha)}=c_i\frac{r_{i,\alpha}^{(t)}}{\displaystyle\sum_{\beta\in I_i}r_{i,\beta}^{(t)}},
\label{eq:normalized_target}
\end{equation}
which is followed by the damped update
\begin{equation}
\omega_{i,t+1}^{(\alpha)}=(1-\eta)\omega_{i,t}^{(\alpha)}+\eta\widetilde{\omega}_{i,t}^{(\alpha)},\qquad\eta=\frac{1}{2}.
\label{eq:damped_update}
\end{equation}
All targets are computed from the same iteration-$t$ coefficients and then applied simultaneously. The normalized targets sum to $c_i$, so the damped update preserves reconstruction when initialized from a valid decomposition. For a term with multiplicity one, its sole coefficient remains fixed to $c_i$. For all the coefficient heuristics, we employed three damped coefficient updates.

The RSD and RV rules used are
\begin{align}
\text{RSD} && r_{i,\alpha}^{(t)}
&=
\sqrt{B_{i,\alpha}^{(t)}+\delta},
\label{eq:binmopt-relative-std}\\
\text{RV}  &&  r_{i,\alpha}^{(t)}
&=
B_{i,\alpha}^{(t)}+\delta,
\label{eq:binmopt-relative-var}
\end{align}
with $\delta=10^{-12}\,\Eh^2$, a regularizer introduced for numerical stability that keeps the normalization well defined when backgrounds vanish. The RSD weighting follows from the local standard-deviation expansion discussed in Appendix~\ref{app:weights}, and it neglects the covariance between the inserted term and its background. RSD assigns larger coefficient magnitudes to groups with larger background standard deviations. RV places a stronger preference on large-background groups through variance-proportional weighting.

For SM, let $m_i=|I_i|$ and define the trial share,
\begin{equation}
q_i=\frac{c_i}{m_i}.
\end{equation}
The change in fragment standard deviation associated with the insertion of $q_iP_i$ is
\begin{equation}
\Delta S_{i,\alpha}^{(t)}
=
\sqrt{
B_{i,\alpha}^{(t)}
+2q_i b_{i,\alpha}^{(t)}
+q_i^2 C_{ii}
}
-
\sqrt{B_{i,\alpha}^{(t)}}.
\label{eq:score_increment}
\end{equation}
This is a trial insertion score relative to the background, not the change in the complete objective under the simultaneous coefficient update. The SM weights are
\begin{equation}
r_{i,\alpha}^{(t)}
=
\exp\!\left[-\beta\Delta S_{i,\alpha}^{(t)}\right],
\qquad
\beta=5\,\Eh^{-1},
\label{eq:binmopt-softmax}
\end{equation}
which are followed by the damped coefficient updates. The covariance term $b_{i,\alpha}^{(t)}$ lets SM account for the interaction between the inserted Pauli word and the remaining fragment. Finally, for all heuristics, we compare the resulting decomposition with the transported initializer using Eq.~\eqref{eq:OverlappingObjective} and retain the lower-cost one. We then refine a shortlist of the lowest-cost candidate covers using fixed-support ICS and append their refined coefficient profiles while retaining the earlier profiles. These refined covers also remain eligible for the final solution.

The O-clique MILP combines these profiles while preserving exact coefficient reconstruction. It is convenient to describe the model using unscaled physical profiles $\mathbf{a}^{(q,r)}$. Let $\mu_{qr}\geq0$ be the continuous multiplier of profile $r$ for clique $q$. The reconstructed coefficient vector on candidate $\mathcal{C}_q$ is
\begin{equation}
\boldsymbol{\omega}^{(q)}
=
\sum_{r\in\mathcal{P}_q}
\mu_{qr}\mathbf{a}^{(q,r)}.
\label{eq:oclique_decoded_coefficients}
\end{equation}
A multiplier scales the complete profile vector, coupling the coefficients of its constituent Pauli words. Several profiles on the same clique may contribute simultaneously. The profile model uses the dimensionless scale parameter $\kappa=2$ and the activation constraints
\begin{equation}
\sum_{r\in\mathcal{P}_q}\mu_{qr}
\leq
\kappa y_q,
\qquad
q=1,\ldots,N_C.
\label{eq:oclique_profile_activation}
\end{equation}

The activation constraint links clique selection to the coefficient profiles that can contribute to its fragment. Since $\mu_{qr}\geq0$, setting $y_q=0$ forces every multiplier on clique $\mathcal{C}_q$ to vanish, giving $\boldsymbol{\omega}^{(q)}=\mathbf{0}$. Setting $y_q=1$ permits a combination of its profiles with total multiplier weight at most $\kappa$. Several profiles may contribute simultaneously to the same selected clique, and their combined coefficient vector defines a single measurement fragment. Component-wise, Eq.~\eqref{eq:oclique_decoded_coefficients} gives
\begin{equation}
\omega_i^{(q)}=\sum_{r\in\mathcal{P}_q}\mu_{qr}a_i^{(q,r)}.
\label{eq:oclique_profile_component}
\end{equation}
The continuous decision variables are therefore the profile multipliers. Changing $\mu_{qr}$ scales every coefficient in profile $\mathbf{a}^{(q,r)}$ by the same factor, while combining several profiles provides additional freedom to redistribute the coefficients within the clique.

The parameter $\kappa$ controls the permitted total weight of the profiles on each selected clique. At $\kappa=1$, two profiles can form a convex combination $\lambda\mathbf{a}^{(q,1)}+(1-\lambda)\mathbf{a}^{(q,2)}$, with $0\leq\lambda\leq1$; the inequality also permits combinations with total weight below one. We use $\kappa=2$ to enlarge this range. For example, $\mu_{q1}=\mu_{q2}=1$ permits the sum $\mathbf{a}^{(q,1)}+\mathbf{a}^{(q,2)}$, and a single profile can receive a multiplier as large as two. The bound constrains the sum of multiplier values, not the number of contributing profiles. These choices describe the permitted combinations within an individual clique; the coefficients across all selected cliques must still satisfy Hamiltonian reconstruction.

Because the profiles are defined as full $N_P$-dimensional vectors with zero entries outside their associated clique, Hamiltonian reconstruction is imposed through,
\begin{equation}
\sum_{q=1}^{N_C}
\sum_{r\in\mathcal{P}_q}
\mu_{qr}a_i^{(q,r)}
=
c_i,
\quad \text{where} \quad
 i=1,\ldots,N_P.
\label{eq:oclique_reconstruction}
\end{equation}
The O-clique optimization is therefore,
\begin{equation}
\begin{aligned}
\underset{\{y_q,\mu_{qr}\}}{\operatorname{minimize}}\quad
& U
=
\sum_{q=1}^{N_C}
\sum_{r\in\mathcal{P}_q}
\sigma_{qr}\mu_{qr},\\[0.3em]
\text{subject to}\quad
& \sum_{r\in\mathcal{P}_q}\mu_{qr}
\leq
\kappa \, y_q, \, q=1,\ldots,N_C \,\text{and}\, \kappa=2,\\
& \sum_{q=1}^{N_C}y_q
\leq K, \quad y_q\in\{0,1\},\\
& \sum_{q=1}^{N_C}
  \sum_{r\in\mathcal{P}_q}
  \mu_{qr}a_i^{(q,r)}
= c_i,\\
& \mu_{qr}\geq0, \quad r\in\mathcal{P}_q.
\end{aligned}
\label{eq:oclique_milp}
\end{equation}
All profile entries and standard deviations are fixed inputs to this model, making it linear, with $N_C$ binary variables and $\sum_qN_q^{\mathrm{prof}}$ continuous variables. Reconstruction ensures coverage of every nonzero Hamiltonian coefficient, and candidate construction ensures compatibility within each selected group. We solve the profile MILP using SCIP through PySCIPOpt \cite{SCIP,hojny2025scipoptimizationsuite100}.

\begin{table*}[t]
\centering
\small
\setlength{\tabcolsep}{7pt}
\renewcommand{\arraystretch}{1.15}

\caption{Measurement objectives $\varepsilon^2 M$ (in $E_{\rm h}^2$) for non-overlapping initializations and after five ICS iterations. $N_P$ denotes the number of Pauli words in the Hamiltonian, excluding the identity term, while values in parentheses give the number of groups $N_G$.
For the post-ICS results, $R=\varepsilon^2M(\mathrm{SI})/\varepsilon^2M(\mathrm{Method})$. Groupings were obtained using CISD covariance estimates during optimization; all reported objectives were subsequently reevaluated using exact FCI covariances for the resulting groupings.}
\label{tab:nonoverlap_cisd}

\begin{tabular}{l r cc cc cc}
\toprule
\multirow{2}{*}{\textbf{System}}
& \multirow{2}{*}{$\boldsymbol{N_P}$}
& \multicolumn{2}{c}{\textbf{Initialization}}
& \multicolumn{2}{c}{\textbf{After ICS}}
& \multicolumn{2}{c}{\textbf{Improvement} $R$} \\
\cmidrule(lr){3-4}
\cmidrule(lr){5-6}
\cmidrule(l){7-8}

& &
\textbf{SI}
& \textbf{MILP}
& \textbf{SI--ICS}
& \textbf{MILP--ICS}
& \textbf{SI--ICS}
& \textbf{MILP--ICS} \\
\midrule

\ce{H4}   & 184  & 1.015 (9)  & \textbf{0.593 (9)}  & 0.706 & \textbf{0.480} & 1.44 & \textbf{2.12} \\
\ce{LiH}  & 275  & 0.276 (20) & \textbf{0.123 (18)} & 0.122 & \textbf{0.080} & 2.26 & \textbf{3.46} \\
\ce{BeH2} & 326  & 0.614 (18) & \textbf{0.356 (17)} & 0.305 & \textbf{0.244} & 2.01 & \textbf{2.51} \\
\ce{H2O}  & 550  & 2.783 (30) & \textbf{1.358 (30)} & 1.025 & \textbf{0.851} & 2.72 & \textbf{3.27} \\
\ce{N2}   & 824  & 3.782 (26) & \textbf{2.109 (28)} & 1.719 & \textbf{1.464} & 2.20 & \textbf{2.58} \\
\ce{H6}   & 918  & 4.010 (35) & \textbf{2.155 (39)} & 1.486 & \textbf{1.354} & 2.70 & \textbf{2.96} \\
\ce{NH3}  & 1393 & 7.768 (62) & \textbf{2.654 (76)} & 2.259 & \textbf{1.553} & 3.44 & \textbf{5.00} \\
\ce{MgO}  & 1616 & 8.745 (53) & \textbf{2.780 (72)} & 1.634 & \textbf{1.039} & 5.35 & \textbf{8.42} \\

\bottomrule
\end{tabular}
\end{table*}

The linear objective $U$ bounds the sum of standard deviations of the reconstructed fragments. For covariance data obtained from the fixed state $\psi$, each clique block $C_q$ is positive semidefinite because $\mathbf z^TC_q\mathbf z=\Var_\psi(\sum_{i\in\mathcal{C}_q}z_iP_i)\geq0$. It therefore defines the seminorm
\begin{equation}
\lVert\mathbf z\rVert_{C_q}=\sqrt{\mathbf z^TC_q\mathbf z},\qquad\mathbf z\in\mathbb{R}^{|\mathcal{C}_q|}.
\end{equation}
Let $\boldsymbol{\omega}_q=(\omega_i^{(q)})_{i\in\mathcal{C}_q}$ denote the restriction of the reconstructed vector to its clique. The triangle inequality and $\mu_{qr}\geq0$ give
\begin{equation}
\begin{aligned}
\sqrt{\boldsymbol{\omega}_q^TC_q\boldsymbol{\omega}_q}&=\left\lVert\sum_{r\in\mathcal{P}_q}\mu_{qr}\mathbf{a}_q^{(r)}\right\rVert_{C_q}\\
&\leq\sum_{r\in\mathcal{P}_q}\mu_{qr}\left\lVert\mathbf{a}_q^{(r)}\right\rVert_{C_q}\\
&=\sum_{r\in\mathcal{P}_q}\mu_{qr}\sigma_{qr}.
\end{aligned}
\label{eq:oclique_triangle_bound}
\end{equation}
Summing over selected candidate cliques, with inactive cliques contributing zero, yields
\begin{equation}
S(X,\omega)=\sum_{q=1}^{N_C}\sqrt{\boldsymbol{\omega}_q^TC_q\boldsymbol{\omega}_q}\leq U,
\label{eq:oclique_upper_bound}
\end{equation}
and hence
\begin{equation}
\epsM(X,\omega)=S(X,\omega)^2\leq U^2.
\label{eq:oclique_upper_bound_squared}
\end{equation}
The equality is achieved when a single nonzero profile contributes to each clique. Combining profiles can make the standard deviation of their sum smaller than the sum of their standard deviations. Thus, $U$ bounds $S$, while $U^2$ bounds the measurement requirement $\epsM$, using the same covariance data. Minimizing $U$ solves the finite-profile surrogate.

After solving Eq.~\eqref{eq:oclique_milp}, we reconstruct the fragment coefficients with the selected cliques and evaluate Eq.~\eqref{eq:OverlappingObjective}.  We then optimize the coefficients using the same procedure as ICS~\cite{Yen2023Deterministic}, holding the selected clique memberships fixed as described in Appendix~\ref{app:weights}. Only terms with multiplicity greater than one have independent coefficient degrees of freedom. This refinement preserves the selected clique memberships and permits coefficients outside the finite profile representation.


\section{Results and discussion}
\label{sec:results}
We use a series of molecular Hamiltonians in the STO-3G basis with frozen cores, mapped via Jordan--Wigner, as a benchmark for FC groups using Tequila \cite{tequila} and PySCF as the backend \cite{PySCF,PySCF2}. The calculations use either exact ground-state covariances from full configuration interaction (FCI)/direct diagonalization of the qubit Hamiltonian or approximate covariances from configuration interaction with single and double excitations (CISD) as indicated. Searches use 12 randomized initial partitions; see Appendix \ref{app:timing} for specific candidate clique budgets and timings for each MILP run. For all molecules, we use an interatomic distance of $1.00$~\AA. For \ce{H4} and \ce{H6}, we consider linear chains. For \ce{H2O}, the H--O--H angle is $107.6^\circ$, and for \ce{NH3}, all H--N--H angles are $107.0^\circ$. All objectives are reported as $\epsM$ in $\Eh^2$. At a target standard error of $1m\Eh$, their numerical values equal the required measurement counts in millions to reach that error. All simulations were run on AMD EPYC 9655 (Zen 5) processors running at 2.7 GHz with 384 MB cache.

\subsection{Binary-optimized non-overlapping groups as ICS initializations and O-clique groupings}
\label{sec:results_nonoverlap}

Table~\ref{tab:nonoverlap_cisd} compares SI and MILP as non-overlapping grouping procedures and as ICS initializations using CISD covariances during the optimization, with the measurement counts reevaluated with the resulting groupings using exact covariances. Our clique-based MILP reduces the non-overlapping measurement requirement relative to SI for all molecules, with reductions ranging from 41.6\% to 68.2\% and an average reduction of 51.8\% relative to SI. The largest reduction occurs for \ce{MgO}, where the measurement requirement decreases from $8.745$ $E_{\rm h}^2$ to $2.780$ $E_{\rm h}^2$. We observe similar results when using exact covariances during optimization, as shown in Table~\ref{tab:nonoverlap} in Appendix~\ref {app:extra_results}. We also considered a hybrid LNS strategy starting from our MILP grouping, which yielded a 0.5\% reduction at a longer search time; we discarded this option as the reduced measurement count does not outweigh the additional costs; see Appendix \ref{app:extra_results}.\\

We find that the non-overlapping groupings produced by our clique-based approach improve initialization for the overlapping ICS technique. We find an average post-ICS reduction of 24\% over the tested molecules. To compare the performance of this initialization against SI-ICS, we use the ratio $R=\varepsilon^2M(\mathrm{SI})/\varepsilon^2M(\mathrm{Method})$. SI-ICS obtains an average ratio of $R=2.76$ with a maximum of 5.35. MILP-ICS outperforms the SI-ICS initialization, obtaining an average of $R=3.79$ and a maximum of 8.42 for \ce{MgO}. This illustrates the benefit of using variance-informed, non-overlapping grouping as initialization for overlapping methods, consistent with previous reports \cite{Choi2023Fluid,huidobromeezs2026discreteflowbasedgenerativemodels,VarSI}. Importantly, the non-overlapping groupings found through the binary optimization outperform the SI-ICS results for \ce{H4} and are comparable for \ce{LiH} and \ce{BeH2} even before coefficient splitting.\\

While variance-informed optimization over non-overlapping groupings clearly reduces the number of measurements, it is important to assess the utility of the resulting groupings for related Hamiltonians to determine whether the optimization cost is worth the measurement reduction. To test the robustness of the MILP-produced grouping, we reuse it for molecular Hamiltonians at different geometries by scanning the potential energy surface (PES). We fix a geometry for the optimization using CISD covariances, then reuse the grouping in new settings. We also consider a refinement variant over the MILP groupings that performs term relocations via a heuristic at different geometries using the new covariances, which we call MILP-R. This avoids reoptimizing the groupings and instead performs a simple refinement over the fixed groups at the optimized geometry $r_{opt}$. Although this refinement is variance-informed, using ICS at the new geometry still requires at least an approximate covariance dictionary. If new terms are found, we use the variance-informed sorted insertion algorithm with the ordered-plus-refinement variant (VarSI-OR) \cite{VarSI}. The VarSI-OR allocation applies only to new terms, allocating them to existing MILP groups, followed by the standard refinement step. For both refinements, we consider 100 refinement sweeps. We evaluate the groupings from $r_{opt}\pm0.5$ \AA, with a 0.1 \AA~step. For \ce{H4} we use $r_{opt}=1.0$ \AA~and $r_{opt}=1.5$ \AA~for the rest of the molecules.\\

\begin{figure}[t!]
    \centering
    \includegraphics[width=0.87\linewidth]{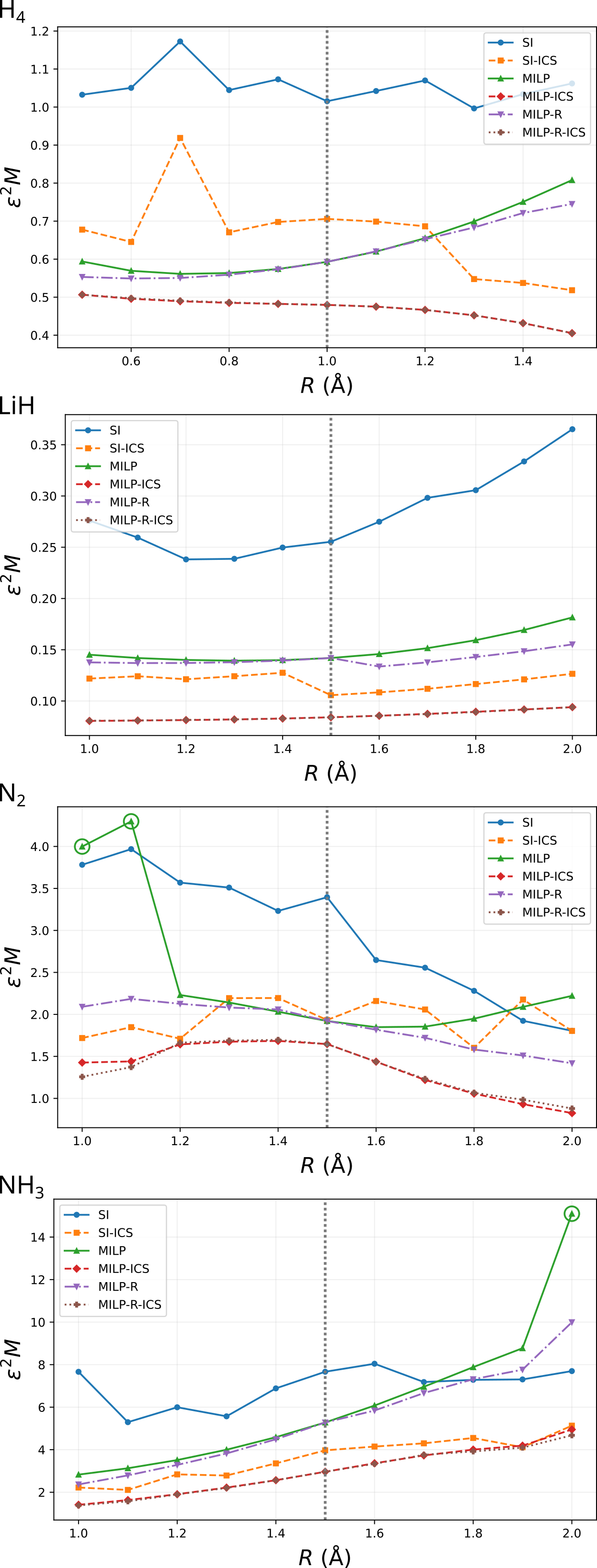}
    \caption{Reuse of MILP-optimized groupings at different geometries. We optimize the MILP grouping at a specific geometry (dashed vertical line) and use that grouping as the measurement scheme for other geometries. Approximate CISD covariance dictionaries used during optimization; results reevaluated for the exact ground state. Geometries with unseen Pauli operators are encircled; for those operators, we use VarSI-OR \cite{VarSI} to assign groups. MILP-R corresponds to the refined variant described in the text.}
    \label{fig:Robustness}
\end{figure}

In Fig. \ref{fig:Robustness} we show the PES scans for \ce{H4}, \ce{LiH}, \ce{N2} and \ce{NH3}. For \ce{H4}, by reusing the grouping, the MILP grouping reduces measurements across the PES scan by 39.7\% compared to SI for the other geometries. This reduction translates to post-ICS initialization with savings of 29.2\% for MILP-ICS versus SI-ICS. We observe that the refinement version, MILP-R, is beneficial when we are far from the optimized geometry. For \ce{LiH}, \ce{N2} and \ce{NH3}, the non-overlapping gains versus SI from MILP/MILP-R are 46.5/49.9\%,  18.6/37.2\% and 11.0/22.2\% respectively. For the overlapping cases MILP-ICS/MILP-R-ICS, these percentages correspond to gains over SI-ICS of 28.2/28.2\%, 30.0/30.3\%, and 16.7/18.1\%, respectively. As expected, when new terms appear at markedly different geometries, MILP grouping performance degrades. The refinement variants used here, and previously proposed in Ref. \cite{VarSI}, can mitigate this effect. This confirms that MILP-optimized groupings at one geometry can serve as initializations at other geometries and can be improved with simple greedy refinement heuristics, while retaining benefits over SI-ICS and demonstrating robustness across PES scans.\\

\paragraph*{Overlapping groupings through O-clique.} 
We next consider overlapping measurement construction using O-clique, in which we select candidate commuting cliques jointly with coefficient profiles before a final ICS refinement. To keep the discrete search space manageable, we cap the candidate pool at 1800 cliques per system. We compare three coefficient-profile guides: relative standard deviation (RSD), relative variance (RV), and softmax scoring (SM), as described in Sec.~\ref{sec:overlap}. Table~\ref{tab:oclique_heuristics} reports the resulting measurement objectives together with their reductions relative to SI-ICS$_{100}$, where the SI initialization is refined using 100 ICS iterations to provide a strongly converged baseline. In contrast, the O-clique solutions are followed by only five ICS iterations on the selected overlapping support, and the table reports the O-clique results from these final five iterations. Approximate CISD covariances are employed during optimization, while all reported measurement objectives are reevaluated using exact FCI covariances. For the systems considered here, the three profile guides generally converge to nearly identical final measurement objectives, and O-clique reduces the measurement requirement relative to SI-ICS$_{100}$ by $16.0\%$--$41.3\%$.

\begin{table}[t]
\centering
\footnotesize
\setlength{\tabcolsep}{5.  pt}
\renewcommand{\arraystretch}{1.15}

\caption{ O-clique measurement objectives for the three coefficient-profile guides compared with SI-ICS after 100 iterations. Each O-clique entry reports $\varepsilon^2M$, with the percentage reduction relative to SI-ICS$_{100}$ shown in parentheses. CISD covariances were used during optimization, while all reported measurement objectives were evaluated using exact FCI covariances. RSD, RV, and SM denote relative standard deviation, relative variance, and softmax-weight heuristics, respectively.}
\label{tab:oclique_heuristics}

\begin{tabular}{lcccc}

\toprule
\multirow{2}{*}{\textbf{System}}
& \multirow{2}{*}{\textbf{SI-ICS$_{100}$}}
& \multicolumn{3}{c}{\textbf{O-clique}} \\
\cmidrule(lr){3-5}
& &
\textbf{RSD}
& \textbf{RV}
& \textbf{SM} \\
\midrule

\ce{H4}
& 0.706
& 0.448 {\scriptsize (37\%)}
& 0.448 {\scriptsize (37\%)}
& 0.448 {\scriptsize (37\%)} \\

\ce{LiH}
& 0.120
& 0.079 {\scriptsize (34\%)}
& 0.079 {\scriptsize (34\%)}
& 0.079 {\scriptsize (34\%)} \\

\ce{BeH2}
& 0.306
& 0.244 {\scriptsize (20\%)}
& 0.244 {\scriptsize (20\%)}
& 0.244 {\scriptsize (20\%)} \\

\ce{H2O}
& 1.029
& 0.835 {\scriptsize (19\%)}
& 0.835 {\scriptsize (19\%)}
& 0.835 {\scriptsize (19\%)} \\

\ce{N2}
& 1.715
& 1.441 {\scriptsize (16\%)}
& 1.441 {\scriptsize (16\%)}
& 1.441 {\scriptsize (16\%)} \\

\ce{H6}
& 1.616
& 1.347 {\scriptsize (17\%)}
& 1.347 {\scriptsize (17\%)}
& 1.347 {\scriptsize (17\%)} \\

\ce{NH3}
& 2.210
& 1.445 {\scriptsize (35\%)}
& 1.445 {\scriptsize (35\%)}
& 1.445 {\scriptsize (35\%)} \\

\ce{MgO}
& 1.535
& 0.901 {\scriptsize (41\%)}
& 0.901 {\scriptsize (41\%)}
& 0.901 {\scriptsize (41\%)} \\

\bottomrule
\end{tabular}
\end{table}

Although O-clique provides only a modest additional reduction relative to the already optimized MILP-ICS baseline, its performance relative to SI-ICS$_{100}$ shows that the method is identifying higher-quality overlapping supports rather than merely benefiting from additional coefficient optimization. Even after 100 ICS iterations, SI-ICS remains consistently above the O-clique solutions, whereas the O-clique results reported here use only five final ICS iterations on the selected support. This indicates that the improvement arises primarily from the choice of overlapping cliques and not from a more extensive coefficient-refinement schedule. At the same time, because the O-clique candidate pool contains cliques derived from the MILP solution, the incremental improvement over MILP-ICS is generally below $10\%$, with a maximum reduction of $13\%$ for \ce{MgO}. A practical strategy is therefore to use MILP-ICS as the primary solution and subsequently run O-clique under a fixed classical time budget, terminating the search when the potential additional measurement reduction no longer justifies the optimization cost.\\

Once the MILP-derived seed is available, the additional O-clique search remains inexpensive for most of the systems considered. As summarized in Table~\ref{tab:oclique_timings} in Appendix \ref{app:timing}, the complete O-clique procedure given the initial MILP guess, including candidate generation, profile construction, optimization, and final coefficient refinement, requires less than five minutes for most calculations. \ce{BeH2} is the notable exception: the optimization reached the imposed time limit without certifying an optimal solution. Among the coefficient-profile guides, RSD and SM generally produce essentially identical measurement objectives with comparable practical runtimes, whereas RV can incur substantially higher costs without a corresponding improvement, as observed for \ce{H2O}. We therefore regard RSD and SM as the most practical profile-generation strategies for O-clique.\\

Overall, these results support a hierarchical measurement-design strategy in which variance-informed MILP first identifies a strong non-overlapping support, MILP-ICS efficiently exploits that support through coefficient splitting, and O-clique is used as a time-budgeted refinement when further reduction of the measurement requirement is valuable. The transferability of the MILP-derived supports across nearby geometries further strengthens this hierarchy, since a grouping optimized at one geometry can remain effective over related Hamiltonians and can therefore amortize its classical construction cost across multiple points on the PES.

\begin{figure*}[t!]
    \centering
    \includegraphics[width=\linewidth]{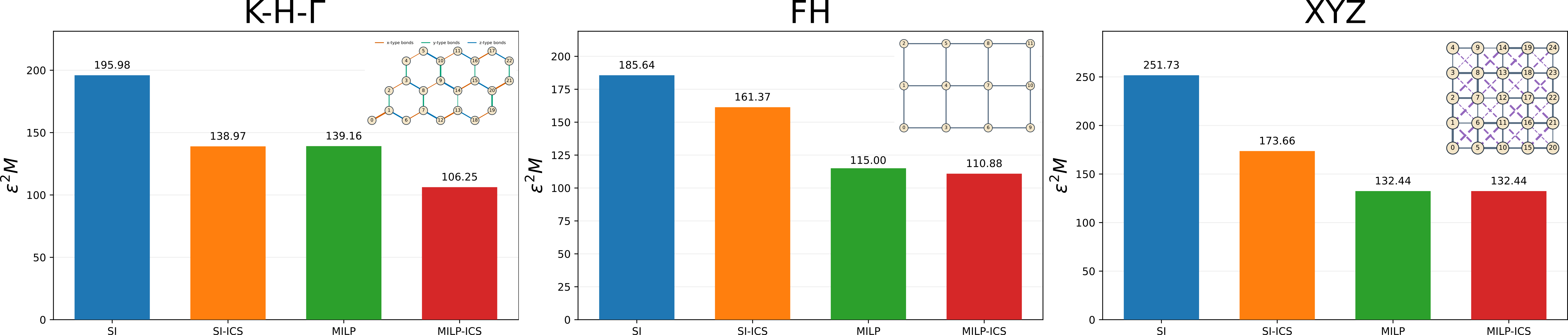}
    \caption{Measurement cost for different lattice Hamiltonians. Comparison of the $\epsM$ values for the measurement groupings found with SI, SI-ICS, MILP, and MILP-ICS. SI-ICS and MILP-ICS use the non-overlapping SI and MILP groups as initialization. We consider a Kitaev honeycomb with anisotropies, a Fermi-Hubbard model, and an XYZ model. Details on the Hamiltonians are available in Appendix \ref{app:Hamiltonians_Lattice}.}
    \label{fig:lattice}
\end{figure*}

\subsection{Performance in Lattice Hamiltonians}
\label{sec:Lattice_results}

To test whether the improvements observed for molecular electronic-structure Hamiltonians transfer to qualitatively different many-body problems, we additionally consider three lattice benchmarks: a spinful Fermi--Hubbard model (FH) at half filling with $N_e=12$ fermions ($N_{\uparrow},N_{\downarrow}=6$), a disordered Kitaev--Heisenberg--$\Gamma$ honeycomb model with Heisenberg and symmetric off-diagonal exchange (K--H--$\Gamma$), and a disordered XYZ model with next-nearest-neighbor interactions. These models draw on paradigmatic settings for strongly correlated fermions, bond-directional and frustrated magnetism, anisotropic exchange, and disorder~\cite{Arovas2022Hubbard,Matsuda2025Kitaev,Wang2019KJGamma,Schulz2018XYZ}. Although measurement-reduction strategies based on geometric partitions have previously been studied for lattice Hamiltonians~\cite{LatticeGeometricPartitions}, the Pauli-grouping methods most closely related to the present work have been developed and benchmarked primarily using molecular electronic-structure Hamiltonians~\cite{Crawford2021SortedInsertion, Yen2023Deterministic}. To our knowledge, their systematic application to the lattice families considered here---and in particular to disordered K--H--$\Gamma$ and disordered next-nearest-neighbor XYZ Hamiltonians---has not previously been reported. These systems therefore provide complementary benchmarks for testing whether covariance-informed non-overlapping and overlapping grouping strategies remain effective outside the molecular setting. For the lattice Hamiltonians, we obtained the reference wavefunctions by exact diagonalization of the qubit Hamiltonians, restricted to the chosen particle-number and spin sector for fermionic models. However, approximate covariances can be used, for example, from matrix-product states optimized with DMRG at a controlled bond dimension. The lattice geometries, particle sectors, coupling conventions, disorder realizations, and explicit Hamiltonian generation details, including the randomized generation of parameters for the K--H--$\Gamma$ and XYZ models, are given in Appendix~\ref{app:Hamiltonians_Lattice}.\\

Fig. \ref{fig:lattice} shows the $\epsM$ counts for the K--H--$\Gamma$, FH, and XYZ models considered for the SI, SI-ICS, MILP, and MILP-ICS methods. In all cases, the MILP and MILP-ICS solutions substantially outperform their corresponding SI and SI-ICS baselines. In particular, the MILP solution can even beat SI-ICS for the FH and XYZ models; for XYZ, the resulting grouping is particularly simple: grouping all X-terms in one group, all Y-terms in another, and all Z-terms in the last. We find improvement ratios for MILP-ICS against SI of $R=$1.84, 1.67, 1.90 for the K--H--$\Gamma$, FH, and XYZ models, respectively. The nominal measurement gains to reach 1$m\Eh$ accuracy for MILP-ICS versus SI-ICS are 32.72, 50.49, and 41.22 $\times 10^6$ for K--H--$\Gamma$, FH, and XYZ, respectively, leading to substantial savings in quantum hardware resources.  \\

To understand the differences between the molecular and lattice results, we analyze the SI algorithm. SI is a greedy algorithm designed on the assumption that variances/covariances of Pauli operators can be replaced with their expectation values under the uniform spherical measure. With this approximation, the Variances of each Pauli operator are $\text{Var}(P_i)=1-\frac{1}{2^{n_q}+1}$, and the covariance between any two terms is zero \cite{Crawford2021SortedInsertion}. This makes SI especially efficient for systems that approach such values. The variance and covariance distributions over molecular systems closely approach these assumptions \cite{huidobromeezs2026discreteflowbasedgenerativemodels}, making SI a reasonable starting point for ICS in such scenarios, although strongly correlated systems show larger variations. 

\begin{figure}[]
    \centering
    \begin{minipage}[t]{0.45\textwidth}
        \centering
        \begin{tikzpicture}
            \node[anchor=south west, inner sep=0] (img) at (0,0)
                {\includegraphics[width=\linewidth]{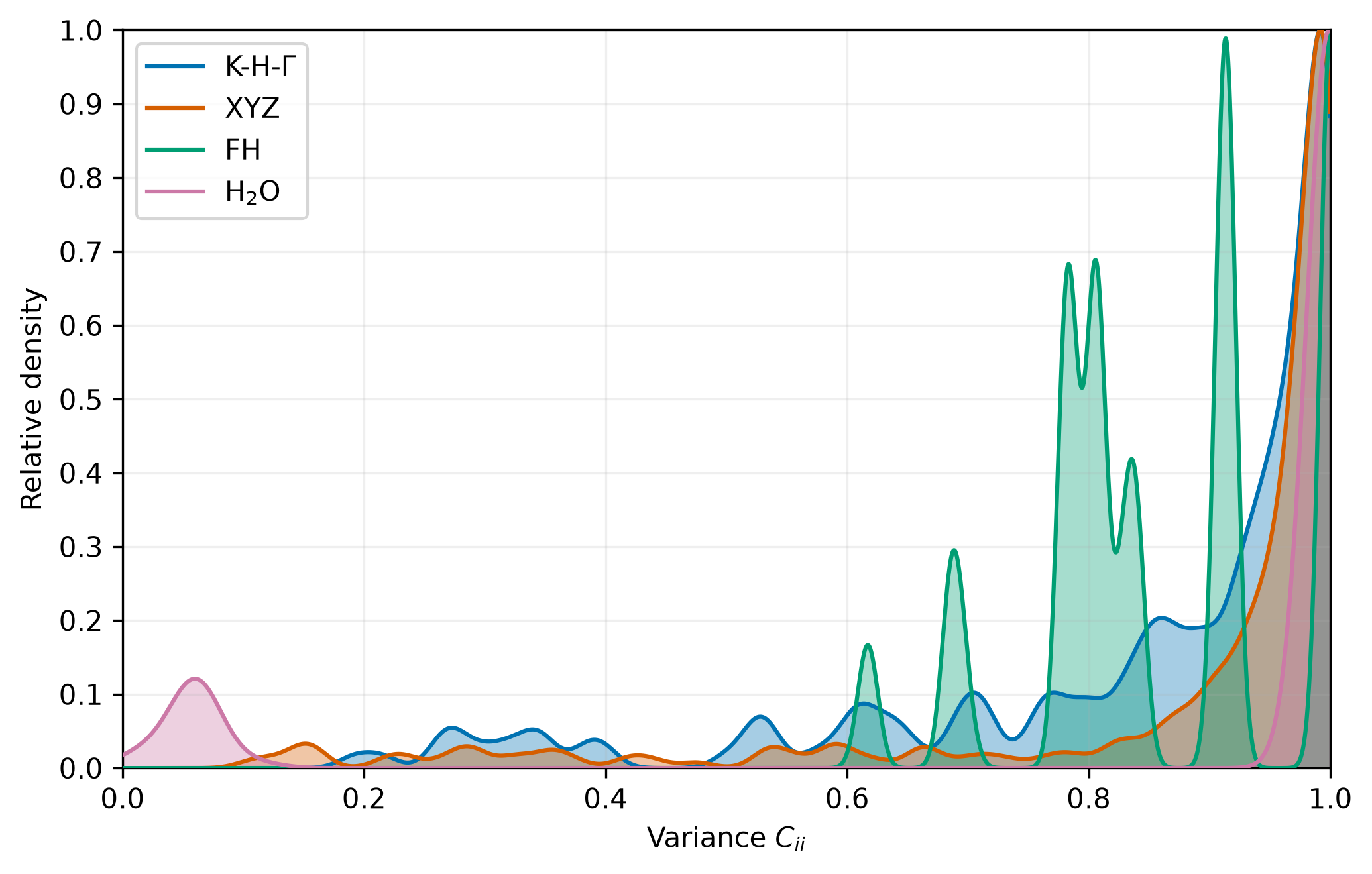}};
            \node[
                anchor=north west,
                fill=white,
                fill opacity=0.85,
                text opacity=1,
                inner sep=0.5pt,
                font=\scriptsize\bfseries
            ] at ([xshift=-10pt,yshift=-8pt]img.north west) {(a)};
        \end{tikzpicture}
    \end{minipage}

    \begin{minipage}[t]{0.45\textwidth}
        \centering
        \begin{tikzpicture}
            \node[anchor=south west, inner sep=0] (img) at (0,0)
                {\includegraphics[width=\linewidth]{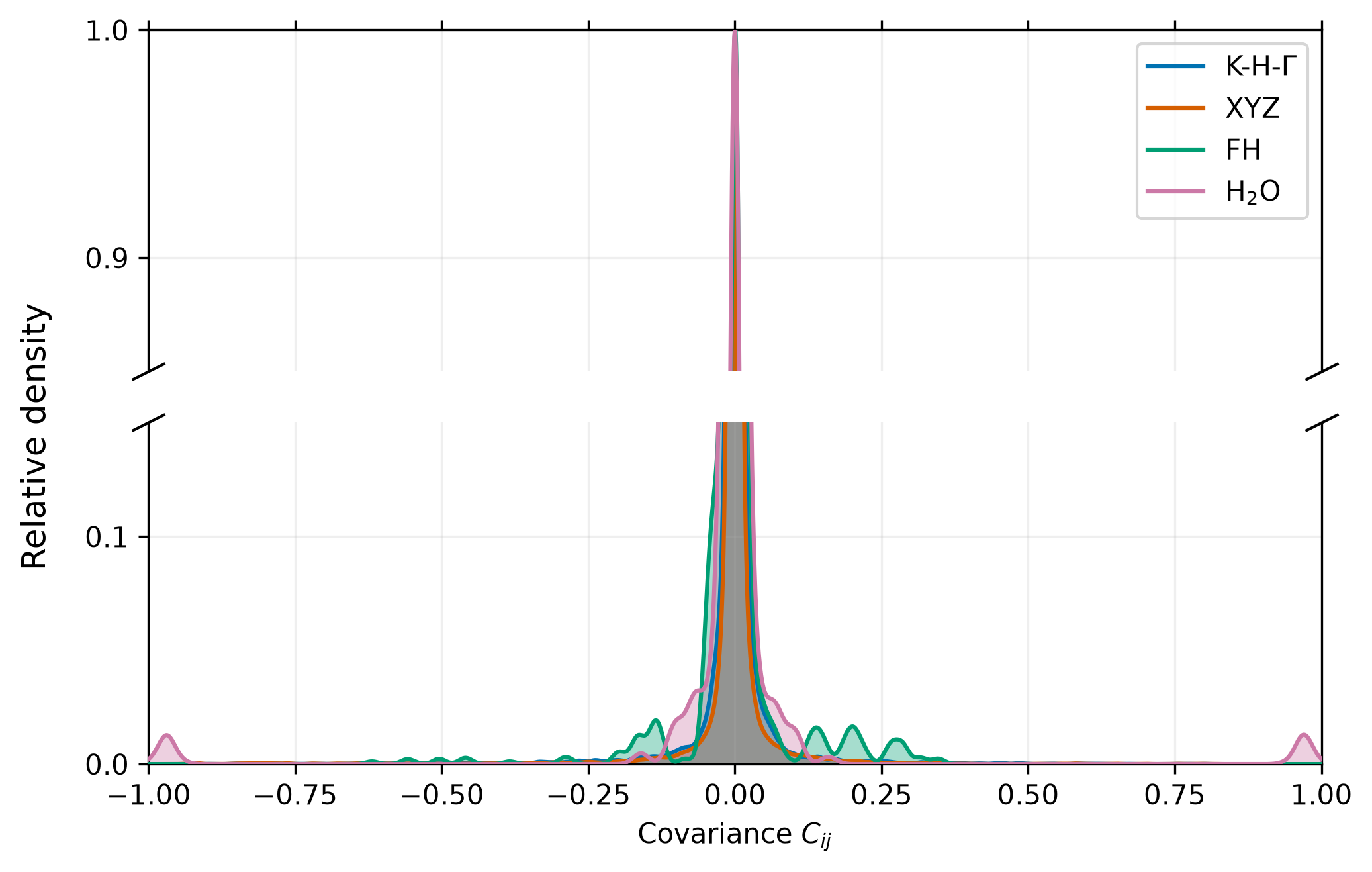}};
            \node[
                anchor=north west,
                fill=white,
                fill opacity=0.85,
                text opacity=1,
                inner sep=0.5pt,
                font=\scriptsize\bfseries
            ] at ([xshift=-10pt,yshift=-8pt]img.north west) {(b)};
        \end{tikzpicture}
    \end{minipage}
    \caption{Distributions of the variances and covariances for the K--H--$\Gamma$, FH, and XYZ model lattice Hamiltonians and \ce{H2O} molecule qubit Hamiltonians. Each density is independently scaled to a maximum of unity. Smoothed distributions were obtained using kernel density estimation with a Gaussian kernel and a smoothing factor of 0.8 for the covariances and of 0.2 for the variances.}
    \label{fig:Covariances}
\end{figure}

For lattice Hamiltonians, the structure might differ significantly from these assumptions, as the cases considered are paradigmatic cases of strong correlation. We show in Fig. \ref{fig:Covariances} the covariance distributions between commuting terms for the model lattice Hamiltonians studied here and for the \ce{H2O} molecule. We independently scale each density to a maximum of unity, allowing comparison of distribution shapes. We observe that while the \ce{H2O} electronic structure Hamiltonian with its ground-state wavefunction produces, in the majority, near-unit variances, the scenario is completely different for the lattice Hamiltonians. This confirms the significant difference between the underlying variance and covariance structures that guide efficient measurement strategies. Although the covariance distribution for these systems falls within the assumptions of SI, the stark difference in variances guides and dominates the measurement strategy, explaining SI's poor performance. Our clique-based binary optimization provides a significantly better non-overlapping initialization as it captures the system's variance and covariance structure, yielding better-informed coefficient splitting and substantial gains even when a non-overlapping grouping is chosen, as shown here.

\section{Conclusions}
\label{sec:conclusions}

Reducing the measurement cost of quantum energy estimation requires optimizing more than the number of commuting groups. The statistical structure within each group determines its contribution to the estimator variance, while the resulting support also constrains the coefficient redistributions available to overlapping measurement strategies. In this work, we have formulated these choices within a classical binary-optimization framework, using a candidate-clique MILP to construct variance-informed non-overlapping groupings and O-clique to investigate the added benefit of optimizing overlapping measurement support directly.

For the molecular benchmarks, the non-overlapping MILP substantially improves over SI. When approximate CISD covariances are used during optimization, MILP reduces $\varepsilon^2M$ by $\sim52\%$ on average relative to SI across the eight molecular systems considered. These improvements persist after coefficient splitting: MILP-ICS reduces the final measurement requirement by an average of $\sim24\%$ relative to SI-ICS. The optimized groupings also remain useful when transferred to geometries other than the one used to construct them, showing that the benefits of the variance-informed partition are not restricted to a single Hamiltonian instance. Simple refinement at the new geometry can further recover performance as the Hamiltonian changes.

Direct optimization of the overlapping support provides a more limited additional benefit for the molecular systems studied here. O-clique reduces the measurement requirement relative to MILP-ICS for every system tested, but the additional reductions of up to $13\%$ require a larger classical search. The three coefficient-profile constructions considered (RSD, RV, and SM) converge to essentially the same final measurement objectives in these benchmarks. Together, these results indicate that optimizing the non-overlapping support is particularly effective: it captures a large fraction of the achievable measurement reduction, provides a strong initialization for subsequent coefficient splitting, and transfers across PES scans. Optimizing the overlapping support directly can further improve the estimator, but the additional gain must be balanced against its classical cost.

The lattice benchmarks show that our conclusions are not specific to molecular electronic-structure Hamiltonians. For the Fermi--Hubbard, disordered Kitaev--Heisenberg--$\Gamma$, and disordered XYZ models considered here, the MILP-based strategies substantially outperform their SI-based counterparts. These systems have interaction and covariance structures that differ qualitatively from those of the molecular benchmarks and therefore provide a complementary test of variance-informed grouping. In particular, the XYZ example shows that binary optimization can recover a compact measurement structure that follows the Hamiltonian's natural operator content. The larger gains observed for these lattice systems also emphasize that grouping strategies developed primarily on molecular Hamiltonians need not retain the same relative performance for other classes of many-body problems.

Several directions follow naturally from these results. The same candidate-generation and MILP framework can be used when only approximate state information is available, as demonstrated here with CISD covariances, and could be combined with other inexpensive covariance estimators for larger systems, making the use of pre-trained wavefunction models from machine learning pipelines to obtain the approximate covariance dictionaries particularly attractive \cite{Mole,ML1RDM,b2026ML2RDM}. The binary representation constructed here can guide training generative models based on graph structures \cite{huidobromeezs2026discreteflowbasedgenerativemodels}, potentially improving search and learning algorithms toward future automatic generation of Hamiltonian partitionings and measurement schemes. Moreover, a fully state-independent variant can be obtained by replacing the covariances by their expectation over pure states uniformly distributed on the unit sphere, as in the criterion underlying SI~\cite{Crawford2021SortedInsertion}. For non-identity Pauli words, this substitution leads to $C_{ij} \mapsto \mathbb{E}_{\psi}[C_{ij}] = \frac{d}{d+1}\delta_{ij}, \quad d=2^{N_q}.$ Such partitions could initialize state-independent overlapping methods such as the recently proposed repacking strategy~\cite{rowland2026overlappedgroupingsRepacking}. Additionally, our clique-based approach can be adapted to approximated Hamiltonians based on hard-core boson approximations of electronic structure Hamiltonians \cite{DavideHCB} and is compatible with recently proposed grouping strategies for measurement reduction in Monte Carlo projective quantum eigensolver \cite{MeasurementMonteCarloProjectiveQE}. Additionally, incorporating symmetry into the measurement design can further reduce the measurement costs \cite{SymmetrySpeedsUp}. Our results show that the covariance structure of the measured state influences measurement-strategy design, making variance-informed methods with approximate covariance dictionaries particularly attractive. More generally, the distinction exposed here between optimizing a non-overlapping support, optimizing coefficients on that support to produce an overlapping grouping, and optimizing the overlapping support itself provides a useful framework for assessing future measurement-reduction strategies.

\section*{Code availability}
The code necessary to reproduce our results is available at \href{https://github.com/ishume94/BinaryMOpt}{https://github.com/ishume94/BinaryMOpt}. 

\begin{acknowledgments}
ILHM acknowledges support from Sandbox AQ's 2025 Research Excellence Scholarship. This research was enabled by support from the Digital Research Alliance of Canada, Compute Ontario, and NSERC Discovery Grant No.~RGPIN-2024-06594.
\end{acknowledgments}

\appendix
\section{Constraint Hamiltonians and Ising representations}
\label{app:ising}
Measurement grouping can be expressed through constraint Hamiltonians whose zero-energy configurations satisfy the assignment and compatibility conditions~\cite{IsingNPProblems,GroupingIsing}. A measurement-dependent objective then ranks the feasible groupings. Although we do not use this approach in the main text because of its scaling with the number of binary variables, we provide the associated constructions using explicit binary-membership variables.

\subsection{Non-overlapping groupings}
\label{sec:clique_constraint}
For non-overlapping groupings, define a Hamiltonian associated with a clique-cover \cite{IsingNPProblems},
\begin{equation}
\small
\begin{split}
H_{\mathrm{no}}={}&A\sum_i\left(1-\sum_{\alpha=1}^{N_f}x_{i,\alpha}\right)^2\\
&+B\sum_{\alpha=1}^{N_f}\Biggl[
\frac12\left(\sum_i x_{i,\alpha}-1\right)\sum_i x_{i,\alpha} -\sum_{(i,j)\in E}x_{i,\alpha}x_{j,\alpha}\Biggr],
\end{split}
\label{eq:H_orig}
\end{equation}
where $A,B>0$. The first term assigns every Pauli word to exactly one group. In the second term, the first sum counts all pairs in a group, and the edge sum subtracts the compatible pairs. Using $x_{i,\alpha}^2=x_{i,\alpha}$ gives
\begin{equation}
\frac12\left[\left(\sum_i x_{i,\alpha}\right)^2-\sum_i x_{i,\alpha}\right]
=\sum_{i<j}x_{i,\alpha}x_{j,\alpha}.
\label{eq:all_pair_identity}
\end{equation}
Consequently,
\begin{equation}
\small
H_{\mathrm{no}}= A\sum_i\left(1-\sum_\alpha x_{i,\alpha}\right)^2 +B\sum_\alpha\sum_{\substack{i<j\\(i,j)\notin E}}x_{i,\alpha}x_{j,\alpha}.
\label{eq:H_no_reduced}
\end{equation}
Each incompatible pair contributes once to the second term. The zero-energy configurations are therefore precisely the feasible non-overlapping clique partitions within the $N_f$ available group labels.

\subsection{Overlapping groupings}
For overlapping groupings, coefficient reconstruction replaces the single-assignment condition. With maximum multiplicity $m_i^{\max}$, the proposed constraint energy is
\begin{equation}
\begin{split}
H_{\mathrm{ov}}={}&\Lambda_1\sum_i\left(c_i-\sum_\alpha\omega_i^{(\alpha)}x_{i,\alpha}\right)^2\\
&+\Lambda_2\sum_i\max\!\left(0,\sum_\alpha x_{i,\alpha}-m_i^{\max}\right)\\
&+\Lambda_3\sum_\alpha\sum_{\substack{i<j\\(i,j)\notin E}}x_{i,\alpha}x_{j,\alpha},
\end{split}
\label{eq:H_overlapping_constraints}
\end{equation}
with positive penalties. The three terms enforce coefficient reconstruction, an optional multiplicity bound, and clique compatibility, respectively. Under Eq.~\eqref{eq:membership_coefficients}, the contribution $\omega_i^{(\alpha)}x_{i,\alpha}$ equals $\omega_i^{(\alpha)}$. For nonzero $c_i$, reconstruction also ensures that $P_i$ appears in at least one group.

When the coefficients and memberships are optimized jointly, Eq.~\eqref{eq:H_overlapping_constraints} defines a mixed binary--continuous constraint energy. For fixed coefficients, the reconstruction term is quadratic in $X$; a quadratic binary representation of the multiplicity penalty requires an additional encoding. 

To include the measurement cost for a profit-cover, one would need to employ a quadratic proxy. For compatible pairs, define $s_i=\sqrt{v_i}$, $s_{ij}=\sqrt{v_i+v_j+w_{ij}}$, and $R_{ij}=s_i+s_j-s_{ij}$. The pairwise proxy is
\begin{equation}
Q(X)=\sum_{\alpha,i}s_i x_{i,\alpha}
-\sum_\alpha\sum_{(i,j)\in E}R_{ij}x_{i,\alpha}x_{j,\alpha}.
\label{eq:quadratic_proxy}
\end{equation}
which provides a quadratic proxy for the measurement cost. In particular, for a feasible non-overlapping partition with positive-semidefinite clique covariance matrices,
\begin{equation}
Q(X)\leq S(\calG),
\qquad [\max\{0,Q(X)\}]^2\leq\epsM(X),
\label{eq:proxy_bound}
\end{equation}
where $S(\calG)=\sum_\alpha\sqrt{V_{G_\alpha}}$, providing a lower bound on the measurement cost.

\subsection{Ising representation}
The substitution
\begin{equation}
x_{i,\alpha}=\frac{1-Z_{i,\alpha}}2
\label{eq:ising_substitution}
\end{equation}
maps a quadratic binary objective to a two-local Ising Hamiltonian, using
\begin{equation}
x_{i,\alpha}x_{j,\beta}
=\frac14\left(I-Z_{i,\alpha}-Z_{j,\beta}+Z_{i,\alpha}Z_{j,\beta}\right).
\label{eq:ising_pair}
\end{equation}
The operators $Z_{i,\alpha}$ act on optimization spins representing the grouping variables. For example, $Q(X)+H_{\mathrm{no}}(X)$ combines the pairwise measurement proxy defined above with the non-overlapping constraints. If $Q_{\min}$ and $Q_{\max}$ bound $Q$ over the binary domain, the condition $\min\{A,B\}>Q_{\max}-Q_{\min}$ ensures a feasible minimizer whenever a feasible partition exists. Every infeasible assignment incurs a penalty of at least $\min\{A,B\}$, exceeding the largest possible improvement in $Q$.

\section{Candidate generation and neighborhood refinement} \label{app:numerical}
Candidate pools combine graph-coloring partitions, greedy partitions, and additional compatible cliques. Coloring uses saturation-degree, largest-degree-first, smallest-last, and randomized sequential orderings. Greedy construction uses first-fit and covariance-aware insertion with original, coefficient-magnitude, individual-variance, and randomized term orderings. We remove duplicate cliques while retaining a complete feasible partition within $N_f$.

Single-term relocations refine the selected and initial partitions using changes in $\epsM$. These moves can generate additional cliques, and we retain the partition with the lowest evaluated measurement requirement. The restricted MILP optimizes over its specified candidate pool.

We considered a hybrid procedure, which adds large-neighborhood search (LNS) refinement to a MILP-derived initialization, MILP-LNS. A trial removes a subset of assignments and reconstructs a feasible partition. Repair is guided by the pairwise standard-deviation saving
\begin{equation}
R_{ij}=\sqrt{v_i}+\sqrt{v_j}-\sqrt{v_i+v_j+w_{ij}}\geq0.
\label{eq:pair_saving}
\end{equation}
The inequality follows from the triangle inequality for standard deviations. These pairwise savings guide the reconstruction but do not determine acceptance: we replace a candidate only when the full measurement objective decreases. The original candidates remain eligible for final selection. For MILP-LNS, each neighborhood removes
\begin{equation}
k=\min\{N_P,\max\{1,\lceil0.15N_P\rceil\}\}
\label{eq:lns_destroy}
\end{equation}
term assignments. With equal probability, terms are sampled uniformly or with probabilities proportional to $|c_i|+0.05\overline{|c|}+10^{-12}\,\Eh$, where $\overline{|c|}$ is the mean absolute coefficient. Sampling is without replacement. Repair visits terms in decreasing incompatibility-degree order and selects compatible destinations using the largest sum of pair rewards $R_{ij}$, with randomized tie-breaking. A randomized feasible reconstruction supplies the fallback when repair fails. The refinement uses a budget of 2000 iterations and one restart, skips repeated assignments, and accepts improvements larger than $10^{-14}\,\Eh^2$.

For O-clique, the candidate pool retains the original SI and non-overlapping MILP cliques, including nonmaximal cliques, together with their maximal extensions corresponding to the SI-ICS and MILP-ICS supports. We construct additional overlapping covers by maximally extending graph-coloring and first-fit partitions under coefficient-based, degree-based, and randomized priorities. We also generate covers through coverage-preserving exchanges and Kempe-chain exchanges between color classes \cite{KempeChanges,Kempe1879OnTG,van2013complexity}. We obtain additional candidates by inserting a Pauli term, removing all incompatible members of the target clique, and retaining both the resulting clique and its greedy maximal extensions. Duplicate clique supports are removed, and complete covers receive priority within bounded sampling, move, and candidate-pool budgets. We always retain the original SI and MILP covers and their ICS expansions. The overlapping optimization permits at most as many fragments as the MILP seed; complete covers used to initialize coefficient profiles must satisfy this bound.

\begin{table*}[t]
\centering
\small
\setlength{\tabcolsep}{7pt}
\renewcommand{\arraystretch}{1.15}

\caption{Measurement objectives $\varepsilon^2M$ (in $E_{\rm h}^2$) for
non-overlapping initializations and after ICS initialization using exact FCI
covariances during optimization.
$N_P$ denotes the number of Pauli words in each Hamiltonian, excluding the
identity term. For the non-overlapping initialization, values in parentheses
give the number of groups $N_G$.
For the post-ICS results,
$R=\varepsilon^2M(\mathrm{SI})/\varepsilon^2M(\mathrm{Method})$.}
\label{tab:nonoverlap}

\begin{tabular}{l r cc cc cc}
\toprule
\multirow{2}{*}{\textbf{System}}
& \multirow{2}{*}{$\boldsymbol{N_P}$}
& \multicolumn{2}{c}{\textbf{Initialization}}
& \multicolumn{2}{c}{\textbf{After ICS}}
& \multicolumn{2}{c}{\textbf{Improvement} $R$} \\
\cmidrule(lr){3-4}
\cmidrule(lr){5-6}
\cmidrule(l){7-8}

& &
\textbf{SI}
& \textbf{MILP}
& \textbf{SI--ICS}
& \textbf{MILP--ICS}
& \textbf{SI--ICS}
& \textbf{MILP--ICS} \\
\midrule
\ce{H4}
& 184
& 1.015 (9)
& 0.593 (9)
& 0.705
& \textbf{0.480}
& 1.44
& \textbf{2.11} \\

\ce{LiH}
& 275
& 0.276 (20)
& 0.122 (18)
& 0.122
& \textbf{0.080}
& 2.26
& \textbf{3.45} \\

\ce{BeH2}
& 326
& 0.613 (18)
& 0.350 (19)
& 0.301
& \textbf{0.240}
& 2.04
& \textbf{2.55} \\

\ce{H2O}
& 550
& 2.783 (30)
& 1.360 (30)
& 1.005
& \textbf{0.836}
& 2.77
& \textbf{3.33} \\

\ce{N2}
& 824
& 3.782 (26)
& 2.101 (25)
& 1.612
& \textbf{1.433}
& 2.35
& \textbf{2.64} \\

\ce{H6}
& 918
& 3.962 (35)
& 2.238 (37)
& 1.563
& \textbf{1.390}
& 2.53
& \textbf{2.85} \\

\ce{NH3}
& 1393
& 7.712 (62)
& 2.468 (77)
& 2.236
& \textbf{1.416}
& 3.45
& \textbf{5.45} \\

\ce{MgO}
& 1616
& 8.745 (53)
& 2.883 (68)
& 1.577
& \textbf{1.180}
& 5.54
& \textbf{7.41} \\

\bottomrule
\end{tabular}
\end{table*}

\section{Coefficient assignment and refinement}
\label{app:weights}
\subsection{Background-standard-deviation weighting}
For iteration $t$, hold the background fragments fixed and neglect their covariances with $P_i$. Adding coefficient $u_\alpha$ changes the fragment standard deviation by
\begin{equation}
\small
\sqrt{B_{i,\alpha}^{(t)}+u_\alpha^2 C_{ii}}-\sqrt{B_{i,\alpha}^{(t)}} =\frac{u_\alpha^2C_{ii}}{2\sqrt{B_{i,\alpha}^{(t)}}}
+O\!\left(\frac{u_\alpha^4C_{ii}^2}{\bigl(B_{i,\alpha}^{(t)}\bigr)^{3/2}}\right),
\label{eq:rsd_expansion}
\end{equation}
for $B_{i,\alpha}^{(t)}>0$ and $u_\alpha^2C_{ii}\ll B_{i,\alpha}^{(t)}$. When $C_{ii}>0$, minimizing the leading contribution under $\sum_{\alpha\in I_i}u_\alpha=c_i$ gives
\begin{equation}
u_\alpha=c_i\frac{\sqrt{B_{i,\alpha}^{(t)}}}{\sum_{\beta\in I_i}\sqrt{B_{i,\beta}^{(t)}}}.
\label{eq:rsd_stationary}
\end{equation}
This local approximation motivates the background-standard-deviation weights. The regularization and reweighting schedules used for coefficient assignment are specified in Sec.~\ref{sec:overlap}.

\subsection{Iterative coefficient splitting}
For fixed group memberships and positive shot fractions $p_\alpha$, coefficient refinement minimizes
\begin{equation}
\mathcal L(\omega;p)
=\sum_\alpha\frac{(\omega^{(\alpha)})^T C_{G_\alpha}\omega^{(\alpha)}}{p_\alpha},
\label{eq:ics_quadratic}
\end{equation}
subject to coefficient reconstruction. Each vector $\omega^{(\alpha)}$ contains the coefficients assigned to $G_\alpha$. Positive-semidefinite covariance matrices and linear constraints make this a convex quadratic problem. Introducing Lagrange multipliers $\lambda_i$ gives
\begin{equation}
\frac{2}{p_\alpha}\sum_{j\in G_\alpha}C_{ij}\omega_j^{(\alpha)}=\lambda_i,
\qquad i\in G_\alpha.
\label{eq:ics_stationarity}
\end{equation}
We eliminate one coefficient per shared term using reconstruction and solve the resulting linear system by least squares. Updating the shot fractions according to Eq.~\eqref{eq:intro-objective} completes one ICS step~\cite{Yen2023Deterministic}. Unless stated otherwise, all calculations employ five ICS iterations, matching the default iteration count from Tequila~\cite{tequila}.

\section{Additional results} \label{app:extra_results}

Table \ref{tab:nonoverlap} shows the results of MILP and MILP-ICS using exact covariances during optimization. In this setting, MILP reduces the non-overlapping measurement requirement by $51.8\%$ on average relative to SI across the eight systems considered. After coefficient splitting, MILP-ICS retains an average reduction of $23.4\%$ relative to SI-ICS. The corresponding average improvement ratios are $R=2.80$ for SI-ICS and $R=3.72$ for MILP-ICS.

Table \ref{tab:milp_lns} shows the results of the hybrid approach. We observe that the results don't represent a significant improvement over the standard MILP route, making the additional LNS step unnecessary for the benchmark cases employed; thus, we capped the \ce{MgO} search with MILP-LNS at 2500 candidate cliques.

\begin{table}[t]
\centering
\small
\setlength{\tabcolsep}{7pt}
\renewcommand{\arraystretch}{1.15}

\caption{Measurement objectives $\varepsilon^2M$ (in $E_{\rm h}^2$)
obtained with MILP-LNS before and after ICS initialization.
$N_P$ denotes the number of non-identity Pauli words in each Hamiltonian.
Exact covariances were employed during optimization.}
\label{tab:milp_lns}

\begin{tabular}{l r cc}
\toprule
\textbf{System}
& $\boldsymbol{N_P}$
& \textbf{MILP-LNS}
& \textbf{MILP-LNS--ICS} \\
\midrule

\ce{H4}   & 184  & 0.593 & \textbf{0.481} \\
\ce{LiH}  & 275  & 0.123 & \textbf{0.080} \\
\ce{BeH2} & 326  & 0.350 & \textbf{0.240} \\
\ce{H2O}  & 550  & 1.360 & \textbf{0.836} \\
\ce{N2}   & 824  & 1.987 & \textbf{1.332} \\
\ce{H6}   & 918  & 2.269 & \textbf{1.416} \\
\ce{NH3}  & 1393 & 2.481 & \textbf{1.437} \\
\ce{MgO}  & 1616 & 6.738 & \textbf{1.137} \\

\bottomrule
\end{tabular}
\end{table}

\subsection{Molecular benchmark candidate-cliques and timings}
\label{app:timing}

Table~\ref{tab:nonoverlap_times} summarizes the number of candidate cliques and the corresponding MILP and ICS timings for the molecular FCI benchmark suite. The ICS timing corresponds to the ICS process from the MILP initialization. The candidate-generation protocol was allowed a maximum pool size of $N_C^{\max}=40000$ cliques; however, the number $N_C=|\mathcal C|$ reported in the table is the number of distinct candidate cliques the protocol actually generated and supplied to the MILP optimization. The reported search time includes candidate-clique construction and the MILP solve, while excluding molecular Hamiltonian generation, FCI wavefunction preparation, and covariance construction. The ICS time corresponds to the subsequent coefficient optimization applied to the MILP grouping. All timings correspond to individual runs. The total O-clique wall times, considering a clique-cap of 1800, are reported in Table~\ref{tab:oclique_timings}. 

\begin{table}[t]
\label{tab:nonoverlap_times}
\centering
\small
\setlength{\tabcolsep}{7pt}
\renewcommand{\arraystretch}{1.15}

\caption{Candidate-clique pool sizes and timings, reported in seconds, for the non-overlapping
MILP calculations in the molecular FCI benchmark suite.
Candidate generation was limited to $N_C^{\max}=40000$ cliques, while
$N_C=|\mathcal{C}|$ denotes the number of candidate cliques generated and
used in the optimization.}
\label{tab:nonoverlap_times}

\begin{tabular}{l r rr}
\toprule
\multirow{2}{*}{\textbf{System}}
& \multirow{2}{*}{$\boldsymbol{N_C}$}
& \multicolumn{2}{c}{\textbf{Timing (s)}} \\
\cmidrule(lr){3-4}

& &
\textbf{MILP search}
& \textbf{ICS} \\
\midrule

\ce{H4}   & 923   & 2.083    & 0.287  \\
\ce{LiH}  & 1786  & 7.939    & 2.040  \\
\ce{BeH2} & 2022  & 10.528   & 4.144  \\
\ce{H2O}  & 3976  & 54.271   & 6.691  \\
\ce{N2}   & 4900  & 170.821  & 11.070 \\
\ce{H6}   & 6136  & 4840.608 & 12.542 \\
\ce{NH3}  & 10940 & 2336.761 & 51.038 \\
\ce{MgO}  & 11003 & 788.540  & 70.190 \\

\bottomrule
\end{tabular}
\end{table}

\begin{table}[t]
\centering
\small
\setlength{\tabcolsep}{7pt}
\renewcommand{\arraystretch}{1.15}

\caption{Total O-clique wall times, in seconds, for the three coefficient-profile guides. The reported time covers the complete O-clique procedure given the MILP initialization. CISD covariances were used during optimization, while the final measurement objectives were evaluated using exact FCI covariances. RSD, RV, and SM denote relative standard deviation, relative variance, and softmax scoring, respectively.}
\label{tab:oclique_timings}

\begin{tabular}{l rrr}
\toprule
\multirow{2}{*}{\textbf{System}} &
\multicolumn{3}{c}{\textbf{Total O-clique time (s)}} \\
\cmidrule(lr){2-4}

& \textbf{RSD}
& \textbf{RV}
& \textbf{SM} \\
\midrule

\ce{H4}
& 16.303
& 16.630
& 21.104 \\

\ce{LiH}
& 16.206
& 13.731
& 255.576 \\

\ce{H2O}
& 114.661
& 1438.786
& 107.853 \\

\ce{N2}
& 39.155
& 39.454
& 42.398 \\

\ce{H6}
& 44.120
& 44.081
& 47.853 \\

\ce{NH3}
& 189.761
& 165.811
& 144.870 \\

\ce{MgO}
& 186.451
& 190.132
& 192.728 \\

\bottomrule
\end{tabular}
\end{table}

\section{Details on Lattice Hamiltonians} \label{app:Hamiltonians_Lattice}

We consider one fermionic lattice model and two direct spin models, all with open boundary conditions. For the spin Hamiltonians, $X_i$, $Y_i$, and $Z_i$ denote Pauli matrices acting on a single spin-$1/2$ degree of freedom at site $i$; no factor of $1/2$ is introduced through the convention $\bm S=\bm\sigma/2$. For the randomized spin models, a single multiplicative factor is applied to the complete exchange tensor on each bond,
\begin{equation}
g_{ij}=1+W_b u_{ij},
\qquad
u_{ij}\sim\mathcal U[-1,1],
\label{eq:lattice_bond_disorder}
\end{equation}
and the local fields are shifted according to
\begin{equation}
\eta_i^a=W_f v_i^a,
\qquad
v_i^a\sim\mathcal U[-1,1],
\qquad a\in\{x,y,z\}.
\label{eq:lattice_field_disorder}
\end{equation}
The bond and field variables are sampled independently, and the disorder realization used here has seed $0$. The explicit Pauli expansions below correspond to this fixed realization; we round their coefficients to three significant figures. We omit identity terms from the Pauli-word counts used for measurement grouping.

\begin{itemize}[leftmargin=*]

\item \textbf{Fermi--Hubbard model.}\\
We use a spinful Fermi--Hubbard Hamiltonian on an open $4\times3$ rectangular lattice containing $N_s=12$ sites,
\begin{equation}
H_{\mathrm{FH}}
=
-t\sum_{\langle i,j\rangle}\sum_{s\in\{\uparrow,\downarrow\}}
\left(
c_{is}^{\dagger}c_{js}
+
c_{js}^{\dagger}c_{is}
\right)
+
U\sum_i n_{i\uparrow}n_{i\downarrow},
\label{eq:lattice_fermi_hubbard}
\end{equation}
where $n_{is}=c_{is}^{\dagger}c_{is}$. Each lattice site contains one spin-up and one spin-down fermionic orbital and can therefore contain up to two fermions. We set $t=1$ and $U=4$, with $t$ defining the energy scale. The calculation is performed at half filling with $N_e=12$ fermions, $N_{\uparrow}=6$, and $N_{\downarrow}=6$. The spin orbitals are ordered as $(i\uparrow,i\downarrow)$ for each site and mapped to $24$ qubits using the Jordan--Wigner transformation. The mapped Hamiltonian contains $104$ non-identity Pauli words. The exact reference state used for the covariance matrix is obtained in the fixed $(N_{\uparrow}, N_{\downarrow})=(6,6)$ particle-number sector.

\item \textbf{Disordered Kitaev honeycomb model.}\\
The K--H--$\Gamma$ spin model is defined on an open honeycomb fragment with $23$ spin-$1/2$ sites obtained from $4\times3$ parent unit cells, with the final unit cell incomplete. The lattice contains $11$ $x$-type, $8$ $y$-type, and $9$ $z$-type nearest-neighbor bonds. The Hamiltonian is
\begin{equation}
\begin{split}
H_{\mathrm{K}}
={}&
\sum_{a\in\{x,y,z\}}
K_a\sum_{\langle i,j\rangle_a}
g_{ij}\sigma_i^a\sigma_j^a\\
&+
J\sum_{\langle i,j\rangle}
g_{ij}
\left(
X_iX_j+Y_iY_j+Z_iZ_j
\right)\\
&+
\Gamma
\sum_{a\in\{x,y,z\}}
\sum_{\langle i,j\rangle_a}
g_{ij}
\left(
\sigma_i^b\sigma_j^c+
\sigma_i^c\sigma_j^b
\right)\\
&+
\sum_i\sum_{a\in\{x,y,z\}}
\left(h_a+\eta_i^a\right)\sigma_i^a,
\end{split}
\label{eq:lattice_kitaev}
\end{equation}
where $\langle i,j\rangle_a$ denotes a bond of Kitaev type $a$ and $\{a,b,c\}=\{x,y,z\}$. We use
\begin{align}
&(K_x,K_y,K_z)=(-1.00,-0.800,-1.20),\\
&J=0.200,\quad
\Gamma=0.400,
\end{align}
together with $(h_x,h_y,h_z)=(0.200,0.100,0.300)$, $W_b=0.500$, and $W_f=0.200$. There is one qubit per site, and the Hamiltonian contains $209$ non-identity Pauli words. The covariance matrix is evaluated using the exact ground state in the full $2^{23}$-dimensional spin Hilbert space. For reproducibility, the realized Pauli Hamiltonian is

\begingroup
\scriptsize
\sloppy
\noindent
\(H_{\mathrm{K}}^{\mathrm{(real)}} = 0.271\,X_{0}\allowbreak\; -\; 1.15\,X_{0}X_{1}\allowbreak\; -\; 0.00281\,Y_{0}\allowbreak\; +\; 0.289\,Y_{0}Y_{1}\allowbreak\; +\; 0.577\,Y_{0}Z_{1}\allowbreak\; +\; 0.345\,Z_{0}\allowbreak\; +\; 0.577\,Z_{0}Y_{1}\allowbreak\; +\; 0.289\,Z_{0}Z_{1}\allowbreak\; +\; 0.169\,X_{1}\allowbreak\; +\; 0.163\,X_{1}X_{2}\allowbreak\; +\; 0.327\,X_{1}Z_{2}\allowbreak\; +\; 0.244\,X_{1}X_{6}\allowbreak\; +\; 0.489\,X_{1}Y_{6}\allowbreak\; +\; 0.229\,Y_{1}\allowbreak\; -\; 0.490\,Y_{1}Y_{2}\allowbreak\; +\; 0.489\,Y_{1}X_{6}\allowbreak\; +\; 0.244\,Y_{1}Y_{6}\allowbreak\; +\; 0.408\,Z_{1}\allowbreak\; +\; 0.327\,Z_{1}X_{2}\allowbreak\; +\; 0.163\,Z_{1}Z_{2}\allowbreak\; -\; 1.22\,Z_{1}Z_{6}\allowbreak\; +\; 0.224\,X_{2}\allowbreak\; -\; 0.500\,X_{2}X_{3}\allowbreak\; +\; 0.171\,Y_{2}\allowbreak\; +\; 0.125\,Y_{2}Y_{3}\allowbreak\; +\; 0.250\,Y_{2}Z_{3}\allowbreak\; +\; 0.274\,Z_{2}\allowbreak\; +\; 0.250\,Z_{2}Y_{3}\allowbreak\; +\; 0.125\,Z_{2}Z_{3}\allowbreak\; +\; 0.379\,X_{3}\allowbreak\; +\; 0.185\,X_{3}X_{4}\allowbreak\; +\; 0.369\,X_{3}Z_{4}\allowbreak\; +\; 0.230\,X_{3}X_{8}\allowbreak\; +\; 0.459\,X_{3}Y_{8}\allowbreak\; +\; 0.235\,Y_{3}\allowbreak\; -\; 0.554\,Y_{3}Y_{4}\allowbreak\; +\; 0.459\,Y_{3}X_{8}\allowbreak\; +\; 0.230\,Y_{3}Y_{8}\allowbreak\; +\; 0.348\,Z_{3}\allowbreak\; +\; 0.369\,Z_{3}X_{4}\allowbreak\; +\; 0.185\,Z_{3}Z_{4}\allowbreak\; -\; 1.15\,Z_{3}Z_{8}\allowbreak\; +\; 0.0737\,X_{4}\allowbreak\; -\; 0.445\,X_{4}X_{5}\allowbreak\; +\; 0.213\,Y_{4}\allowbreak\; +\; 0.111\,Y_{4}Y_{5}\allowbreak\; +\; 0.223\,Y_{4}Z_{5}\allowbreak\; +\; 0.121\,Z_{4}\allowbreak\; +\; 0.223\,Z_{4}Y_{5}\allowbreak\; +\; 0.111\,Z_{4}Z_{5}\allowbreak\; +\; 0.0141\,X_{5}\allowbreak\; +\; 0.264\,X_{5}X_{10}\allowbreak\; +\; 0.528\,X_{5}Y_{10}\allowbreak\; -\; 0.0255\,Y_{5}\allowbreak\; +\; 0.528\,Y_{5}X_{10}\allowbreak\; +\; 0.264\,Y_{5}Y_{10}\allowbreak\; +\; 0.388\,Z_{5}\allowbreak\; -\; 1.32\,Z_{5}Z_{10}\allowbreak\; +\; 0.127\,X_{6}\allowbreak\; -\; 0.615\,X_{6}X_{7}\allowbreak\; +\; 0.106\,Y_{6}\allowbreak\; +\; 0.154\,Y_{6}Y_{7}\allowbreak\; +\; 0.307\,Y_{6}Z_{7}\allowbreak\; +\; 0.239\,Z_{6}\allowbreak\; +\; 0.307\,Z_{6}Y_{7}\allowbreak\; +\; 0.154\,Z_{6}Z_{7}\allowbreak\; +\; 0.350\,X_{7}\allowbreak\; +\; 0.236\,X_{7}X_{8}\allowbreak\; +\; 0.472\,X_{7}Z_{8}\allowbreak\; +\; 0.271\,X_{7}X_{12}\allowbreak\; +\; 0.542\,X_{7}Y_{12}\allowbreak\; +\; 0.285\,Y_{7}\allowbreak\; -\; 0.708\,Y_{7}Y_{8}\allowbreak\; +\; 0.542\,Y_{7}X_{12}\allowbreak\; +\; 0.271\,Y_{7}Y_{12}\allowbreak\; +\; 0.385\,Z_{7}\allowbreak\; +\; 0.472\,Z_{7}X_{8}\allowbreak\; +\; 0.236\,Z_{7}Z_{8}\allowbreak\; -\; 1.35\,Z_{7}Z_{12}\allowbreak\; +\; 0.0481\,X_{8}\allowbreak\; -\; 0.472\,X_{8}X_{9}\allowbreak\; -\; 0.0192\,Y_{8}\allowbreak\; +\; 0.118\,Y_{8}Y_{9}\allowbreak\; +\; 0.236\,Y_{8}Z_{9}\allowbreak\; +\; 0.366\,Z_{8}\allowbreak\; +\; 0.236\,Z_{8}Y_{9}\allowbreak\; +\; 0.118\,Z_{8}Z_{9}\allowbreak\; +\; 0.124\,X_{9}\allowbreak\; +\; 0.282\,X_{9}X_{10}\allowbreak\; +\; 0.564\,X_{9}Z_{10}\allowbreak\; +\; 0.284\,X_{9}X_{14}\allowbreak\; +\; 0.569\,X_{9}Y_{14}\allowbreak\; +\; 0.135\,Y_{9}\allowbreak\; -\; 0.847\,Y_{9}Y_{10}\allowbreak\; +\; 0.569\,Y_{9}X_{14}\allowbreak\; +\; 0.284\,Y_{9}Y_{14}\allowbreak\; +\; 0.370\,Z_{9}\allowbreak\; +\; 0.564\,Z_{9}X_{10}\allowbreak\; +\; 0.282\,Z_{9}Z_{10}\allowbreak\; -\; 1.42\,Z_{9}Z_{14}\allowbreak\; +\; 0.102\,X_{10}\allowbreak\; -\; 0.493\,X_{10}X_{11}\allowbreak\; +\; 0.222\,Y_{10}\allowbreak\; +\; 0.123\,Y_{10}Y_{11}\allowbreak\; +\; 0.247\,Y_{10}Z_{11}\allowbreak\; +\; 0.301\,Z_{10}\allowbreak\; +\; 0.247\,Z_{10}Y_{11}\allowbreak\; +\; 0.123\,Z_{10}Z_{11}\allowbreak\; +\; 0.387\,X_{11}\allowbreak\; +\; 0.202\,X_{11}X_{16}\allowbreak\; +\; 0.403\,X_{11}Y_{16}\allowbreak\; -\; 0.0959\,Y_{11}\allowbreak\; +\; 0.403\,Y_{11}X_{16}\allowbreak\; +\; 0.202\,Y_{11}Y_{16}\allowbreak\; +\; 0.221\,Z_{11}\allowbreak\; -\; 1.01\,Z_{11}Z_{16}\allowbreak\; +\; 0.208\,X_{12}\allowbreak\; -\; 0.935\,X_{12}X_{13}\allowbreak\; +\; 0.170\,Y_{12}\allowbreak\; +\; 0.234\,Y_{12}Y_{13}\allowbreak\; +\; 0.467\,Y_{12}Z_{13}\allowbreak\; +\; 0.400\,Z_{12}\allowbreak\; +\; 0.467\,Z_{12}Y_{13}\allowbreak\; +\; 0.234\,Z_{12}Z_{13}\allowbreak\; +\; 0.208\,X_{13}\allowbreak\; +\; 0.101\,X_{13}X_{14}\allowbreak\; +\; 0.203\,X_{13}Z_{14}\allowbreak\; +\; 0.189\,X_{13}X_{18}\allowbreak\; +\; 0.377\,X_{13}Y_{18}\allowbreak\; +\; 0.247\,Y_{13}\allowbreak\; -\; 0.304\,Y_{13}Y_{14}\allowbreak\; +\; 0.377\,Y_{13}X_{18}\allowbreak\; +\; 0.189\,Y_{13}Y_{18}\allowbreak\; +\; 0.228\,Z_{13}\allowbreak\; +\; 0.203\,Z_{13}X_{14}\allowbreak\; +\; 0.101\,Z_{13}Z_{14}\allowbreak\; -\; 0.944\,Z_{13}Z_{18}\allowbreak\; +\; 0.217\,X_{14}\allowbreak\; -\; 0.773\,X_{14}X_{15}\allowbreak\; +\; 0.132\,Y_{14}\allowbreak\; +\; 0.193\,Y_{14}Y_{15}\allowbreak\; +\; 0.387\,Y_{14}Z_{15}\allowbreak\; +\; 0.153\,Z_{14}\allowbreak\; +\; 0.387\,Z_{14}Y_{15}\allowbreak\; +\; 0.193\,Z_{14}Z_{15}\allowbreak\; +\; 0.0489\,X_{15}\allowbreak\; +\; 0.140\,X_{15}X_{16}\allowbreak\; +\; 0.280\,X_{15}Z_{16}\allowbreak\; +\; 0.229\,X_{15}X_{20}\allowbreak\; +\; 0.458\,X_{15}Y_{20}\allowbreak\; -\; 0.0446\,Y_{15}\allowbreak\; -\; 0.420\,Y_{15}Y_{16}\allowbreak\; +\; 0.458\,Y_{15}X_{20}\allowbreak\; +\; 0.229\,Y_{15}Y_{20}\allowbreak\; +\; 0.460\,Z_{15}\allowbreak\; +\; 0.280\,Z_{15}X_{16}\allowbreak\; +\; 0.140\,Z_{15}Z_{16}\allowbreak\; -\; 1.14\,Z_{15}Z_{20}\allowbreak\; +\; 0.246\,X_{16}\allowbreak\; -\; 0.780\,X_{16}X_{17}\allowbreak\; +\; 0.106\,Y_{16}\allowbreak\; +\; 0.195\,Y_{16}Y_{17}\allowbreak\; +\; 0.390\,Y_{16}Z_{17}\allowbreak\; +\; 0.200\,Z_{16}\allowbreak\; +\; 0.390\,Z_{16}Y_{17}\allowbreak\; +\; 0.195\,Z_{16}Z_{17}\allowbreak\; +\; 0.0625\,X_{17}\allowbreak\; +\; 0.194\,X_{17}X_{22}\allowbreak\; +\; 0.388\,X_{17}Y_{22}\allowbreak\; +\; 0.225\,Y_{17}\allowbreak\; +\; 0.388\,Y_{17}X_{22}\allowbreak\; +\; 0.194\,Y_{17}Y_{22}\allowbreak\; +\; 0.303\,Z_{17}\allowbreak\; -\; 0.971\,Z_{17}Z_{22}\allowbreak\; +\; 0.0340\,X_{18}\allowbreak\; -\; 0.578\,X_{18}X_{19}\allowbreak\; -\; 0.0832\,Y_{18}\allowbreak\; +\; 0.144\,Y_{18}Y_{19}\allowbreak\; +\; 0.289\,Y_{18}Z_{19}\allowbreak\; +\; 0.370\,Z_{18}\allowbreak\; +\; 0.289\,Z_{18}Y_{19}\allowbreak\; +\; 0.144\,Z_{18}Z_{19}\allowbreak\; +\; 0.111\,X_{19}\allowbreak\; +\; 0.274\,X_{19}X_{20}\allowbreak\; +\; 0.549\,X_{19}Z_{20}\allowbreak\; +\; 0.0296\,Y_{19}\allowbreak\; -\; 0.823\,Y_{19}Y_{20}\allowbreak\; +\; 0.113\,Z_{19}\allowbreak\; +\; 0.549\,Z_{19}X_{20}\allowbreak\; +\; 0.274\,Z_{19}Z_{20}\allowbreak\; +\; 0.137\,X_{20}\allowbreak\; -\; 1.01\,X_{20}X_{21}\allowbreak\; +\; 0.0491\,Y_{20}\allowbreak\; +\; 0.253\,Y_{20}Y_{21}\allowbreak\; +\; 0.505\,Y_{20}Z_{21}\allowbreak\; +\; 0.322\,Z_{20}\allowbreak\; +\; 0.505\,Z_{20}Y_{21}\allowbreak\; +\; 0.253\,Z_{20}Z_{21}\allowbreak\; +\; 0.329\,X_{21}\allowbreak\; +\; 0.172\,X_{21}X_{22}\allowbreak\; +\; 0.345\,X_{21}Z_{22}\allowbreak\; +\; 0.158\,Y_{21}\allowbreak\; -\; 0.517\,Y_{21}Y_{22}\allowbreak\; +\; 0.155\,Z_{21}\allowbreak\; +\; 0.345\,Z_{21}X_{22}\allowbreak\; +\; 0.172\,Z_{21}Z_{22}\allowbreak\; +\; 0.160\,X_{22}\allowbreak\; +\; 0.0415\,Y_{22}\allowbreak\; +\; 0.208\,Z_{22}.\)
\par
\endgroup

\item \textbf{Disordered XYZ model.}\\
We use an open $5\times5$ square lattice with one spin-$1/2$ degree of freedom per site, giving $25$ qubits. The lattice has $40$ nearest-neighbor bonds and $32$ diagonal next-nearest-neighbor bonds. The Hamiltonian is
\begin{equation}
\small
\begin{split}
H_{\mathrm{XYZ}}
={}&
\sum_{\langle i,j\rangle}
g_{ij}
\left(
J_xX_iX_j+J_yY_iY_j+J_zZ_iZ_j
\right)\\
&+
J_2
\sum_{\langle\!\langle i,j\rangle\!\rangle}
g_{ij}
\left(
X_iX_j+Y_iY_j+Z_iZ_j
\right)\\
&+
\sum_i
\left[
(h_x+\eta_i^x)X_i
+
(h_y+\eta_i^y)Y_i
+
(h_z+\eta_i^z)Z_i
\right].
\end{split}
\label{eq:lattice_xyz}
\end{equation}
The parameters are
\begin{equation}
(J_x,J_y,J_z)=(1.00,0.800,1.20),\quad
J_2=0.500,
\end{equation}
with $(h_x,h_y,h_z)=(0.200,0.100,0.300)$, $W_b=0.500$, and $W_f=0.300$. The Hamiltonian contains $291$ nonidentity Pauli words, and the covariance matrix is evaluated using the exact ground state in the full $2^{25}$-dimensional spin Hilbert space. The realized Pauli Hamiltonian is

\begingroup
\scriptsize
\sloppy
\noindent
\(H_{\mathrm{XYZ}}^{\mathrm{(real)}} = 0.306\,X_{0}\allowbreak\; +\; 1.44\,X_{0}X_{1}\allowbreak\; +\; 0.816\,X_{0}X_{5}\allowbreak\; +\; 0.710\,X_{0}X_{6}\allowbreak\; -\; 0.0542\,Y_{0}\allowbreak\; +\; 1.15\,Y_{0}Y_{1}\allowbreak\; +\; 0.653\,Y_{0}Y_{5}\allowbreak\; +\; 0.710\,Y_{0}Y_{6}\allowbreak\; +\; 0.367\,Z_{0}\allowbreak\; +\; 1.73\,Z_{0}Z_{1}\allowbreak\; +\; 0.980\,Z_{0}Z_{5}\allowbreak\; +\; 0.710\,Z_{0}Z_{6}\allowbreak\; +\; 0.154\,X_{1}\allowbreak\; +\; 1.22\,X_{1}X_{2}\allowbreak\; +\; 0.630\,X_{1}X_{5}\allowbreak\; +\; 0.626\,X_{1}X_{6}\allowbreak\; +\; 0.721\,X_{1}X_{7}\allowbreak\; +\; 0.294\,Y_{1}\allowbreak\; +\; 0.978\,Y_{1}Y_{2}\allowbreak\; +\; 0.630\,Y_{1}Y_{5}\allowbreak\; +\; 0.500\,Y_{1}Y_{6}\allowbreak\; +\; 0.721\,Y_{1}Y_{7}\allowbreak\; +\; 0.462\,Z_{1}\allowbreak\; +\; 1.47\,Z_{1}Z_{2}\allowbreak\; +\; 0.630\,Z_{1}Z_{5}\allowbreak\; +\; 0.751\,Z_{1}Z_{6}\allowbreak\; +\; 0.721\,Z_{1}Z_{7}\allowbreak\; +\; 0.236\,X_{2}\allowbreak\; +\; 0.923\,X_{2}X_{3}\allowbreak\; +\; 0.292\,X_{2}X_{6}\allowbreak\; +\; 1.15\,X_{2}X_{7}\allowbreak\; +\; 0.658\,X_{2}X_{8}\allowbreak\; +\; 0.207\,Y_{2}\allowbreak\; +\; 0.738\,Y_{2}Y_{3}\allowbreak\; +\; 0.292\,Y_{2}Y_{6}\allowbreak\; +\; 0.918\,Y_{2}Y_{7}\allowbreak\; +\; 0.658\,Y_{2}Y_{8}\allowbreak\; +\; 0.261\,Z_{2}\allowbreak\; +\; 1.11\,Z_{2}Z_{3}\allowbreak\; +\; 0.292\,Z_{2}Z_{6}\allowbreak\; +\; 1.38\,Z_{2}Z_{7}\allowbreak\; +\; 0.658\,Z_{2}Z_{8}\allowbreak\; +\; 0.468\,X_{3}\allowbreak\; +\; 0.557\,X_{3}X_{4}\allowbreak\; +\; 0.353\,X_{3}X_{7}\allowbreak\; +\; 1.32\,X_{3}X_{8}\allowbreak\; +\; 0.258\,X_{3}X_{9}\allowbreak\; +\; 0.303\,Y_{3}\allowbreak\; +\; 0.445\,Y_{3}Y_{4}\allowbreak\; +\; 0.353\,Y_{3}Y_{7}\allowbreak\; +\; 1.06\,Y_{3}Y_{8}\allowbreak\; +\; 0.258\,Y_{3}Y_{9}\allowbreak\; +\; 0.371\,Z_{3}\allowbreak\; +\; 0.668\,Z_{3}Z_{4}\allowbreak\; +\; 0.353\,Z_{3}Z_{7}\allowbreak\; +\; 1.58\,Z_{3}Z_{8}\allowbreak\; +\; 0.258\,Z_{3}Z_{9}\allowbreak\; +\; 0.0106\,X_{4}\allowbreak\; +\; 0.374\,X_{4}X_{8}\allowbreak\; +\; 0.769\,X_{4}X_{9}\allowbreak\; +\; 0.269\,Y_{4}\allowbreak\; +\; 0.374\,Y_{4}Y_{8}\allowbreak\; +\; 0.615\,Y_{4}Y_{9}\allowbreak\; +\; 0.0319\,Z_{4}\allowbreak\; +\; 0.374\,Z_{4}Z_{8}\allowbreak\; +\; 0.922\,Z_{4}Z_{9}\allowbreak\; -\; 0.0789\,X_{5}\allowbreak\; +\; 1.18\,X_{5}X_{6}\allowbreak\; +\; 1.35\,X_{5}X_{10}\allowbreak\; +\; 0.719\,X_{5}X_{11}\allowbreak\; -\; 0.0882\,Y_{5}\allowbreak\; +\; 0.943\,Y_{5}Y_{6}\allowbreak\; +\; 1.08\,Y_{5}Y_{10}\allowbreak\; +\; 0.719\,Y_{5}Y_{11}\allowbreak\; +\; 0.433\,Z_{5}\allowbreak\; +\; 1.42\,Z_{5}Z_{6}\allowbreak\; +\; 1.63\,Z_{5}Z_{10}\allowbreak\; +\; 0.719\,Z_{5}Z_{11}\allowbreak\; +\; 0.0905\,X_{6}\allowbreak\; +\; 0.590\,X_{6}X_{7}\allowbreak\; +\; 0.562\,X_{6}X_{10}\allowbreak\; +\; 1.41\,X_{6}X_{11}\allowbreak\; +\; 0.340\,X_{6}X_{12}\allowbreak\; +\; 0.109\,Y_{6}\allowbreak\; +\; 0.472\,Y_{6}Y_{7}\allowbreak\; +\; 0.562\,Y_{6}Y_{10}\allowbreak\; +\; 1.13\,Y_{6}Y_{11}\allowbreak\; +\; 0.340\,Y_{6}Y_{12}\allowbreak\; +\; 0.208\,Z_{6}\allowbreak\; +\; 0.708\,Z_{6}Z_{7}\allowbreak\; +\; 0.562\,Z_{6}Z_{10}\allowbreak\; +\; 1.69\,Z_{6}Z_{11}\allowbreak\; +\; 0.340\,Z_{6}Z_{12}\allowbreak\; +\; 0.424\,X_{7}\allowbreak\; +\; 1.42\,X_{7}X_{8}\allowbreak\; +\; 0.462\,X_{7}X_{11}\allowbreak\; +\; 0.617\,X_{7}X_{12}\allowbreak\; +\; 0.445\,X_{7}X_{13}\allowbreak\; +\; 0.377\,Y_{7}\allowbreak\; +\; 1.14\,Y_{7}Y_{8}\allowbreak\; +\; 0.462\,Y_{7}Y_{11}\allowbreak\; +\; 0.493\,Y_{7}Y_{12}\allowbreak\; +\; 0.445\,Y_{7}Y_{13}\allowbreak\; +\; 0.428\,Z_{7}\allowbreak\; +\; 1.71\,Z_{7}Z_{8}\allowbreak\; +\; 0.462\,Z_{7}Z_{11}\allowbreak\; +\; 0.740\,Z_{7}Z_{12}\allowbreak\; +\; 0.445\,Z_{7}Z_{13}\allowbreak\; -\; 0.0279\,X_{8}\allowbreak\; +\; 1.01\,X_{8}X_{9}\allowbreak\; +\; 0.615\,X_{8}X_{12}\allowbreak\; +\; 1.17\,X_{8}X_{13}\allowbreak\; +\; 0.696\,X_{8}X_{14}\allowbreak\; -\; 0.0789\,Y_{8}\allowbreak\; +\; 0.807\,Y_{8}Y_{9}\allowbreak\; +\; 0.615\,Y_{8}Y_{12}\allowbreak\; +\; 0.935\,Y_{8}Y_{13}\allowbreak\; +\; 0.696\,Y_{8}Y_{14}\allowbreak\; +\; 0.399\,Z_{8}\allowbreak\; +\; 1.21\,Z_{8}Z_{9}\allowbreak\; +\; 0.615\,Z_{8}Z_{12}\allowbreak\; +\; 1.40\,Z_{8}Z_{13}\allowbreak\; +\; 0.696\,Z_{8}Z_{14}\allowbreak\; +\; 0.0855\,X_{9}\allowbreak\; +\; 0.561\,X_{9}X_{13}\allowbreak\; +\; 0.507\,X_{9}X_{14}\allowbreak\; +\; 0.153\,Y_{9}\allowbreak\; +\; 0.561\,Y_{9}Y_{13}\allowbreak\; +\; 0.406\,Y_{9}Y_{14}\allowbreak\; +\; 0.405\,Z_{9}\allowbreak\; +\; 0.561\,Z_{9}Z_{13}\allowbreak\; +\; 0.608\,Z_{9}Z_{14}\allowbreak\; +\; 0.0530\,X_{10}\allowbreak\; +\; 0.944\,X_{10}X_{11}\allowbreak\; +\; 0.966\,X_{10}X_{15}\allowbreak\; +\; 0.405\,X_{10}X_{16}\allowbreak\; +\; 0.283\,Y_{10}\allowbreak\; +\; 0.755\,Y_{10}Y_{11}\allowbreak\; +\; 0.773\,Y_{10}Y_{15}\allowbreak\; +\; 0.405\,Y_{10}Y_{16}\allowbreak\; +\; 0.302\,Z_{10}\allowbreak\; +\; 1.13\,Z_{10}Z_{11}\allowbreak\; +\; 1.16\,Z_{10}Z_{15}\allowbreak\; +\; 0.405\,Z_{10}Z_{16}\allowbreak\; +\; 0.481\,X_{11}\allowbreak\; +\; 0.700\,X_{11}X_{12}\allowbreak\; +\; 0.576\,X_{11}X_{15}\allowbreak\; +\; 1.14\,X_{11}X_{16}\allowbreak\; +\; 0.624\,X_{11}X_{17}\allowbreak\; -\; 0.194\,Y_{11}\allowbreak\; +\; 0.560\,Y_{11}Y_{12}\allowbreak\; +\; 0.576\,Y_{11}Y_{15}\allowbreak\; +\; 0.916\,Y_{11}Y_{16}\allowbreak\; +\; 0.624\,Y_{11}Y_{17}\allowbreak\; +\; 0.182\,Z_{11}\allowbreak\; +\; 0.840\,Z_{11}Z_{12}\allowbreak\; +\; 0.576\,Z_{11}Z_{15}\allowbreak\; +\; 1.37\,Z_{11}Z_{16}\allowbreak\; +\; 0.624\,Z_{11}Z_{17}\allowbreak\; +\; 0.212\,X_{12}\allowbreak\; +\; 0.975\,X_{12}X_{13}\allowbreak\; +\; 0.659\,X_{12}X_{16}\allowbreak\; +\; 0.971\,X_{12}X_{17}\allowbreak\; +\; 0.263\,X_{12}X_{18}\allowbreak\; +\; 0.205\,Y_{12}\allowbreak\; +\; 0.780\,Y_{12}Y_{13}\allowbreak\; +\; 0.659\,Y_{12}Y_{16}\allowbreak\; +\; 0.776\,Y_{12}Y_{17}\allowbreak\; +\; 0.263\,Y_{12}Y_{18}\allowbreak\; +\; 0.450\,Z_{12}\allowbreak\; +\; 1.17\,Z_{12}Z_{13}\allowbreak\; +\; 0.659\,Z_{12}Z_{16}\allowbreak\; +\; 1.16\,Z_{12}Z_{17}\allowbreak\; +\; 0.263\,Z_{12}Z_{18}\allowbreak\; +\; 0.212\,X_{13}\allowbreak\; +\; 0.722\,X_{13}X_{14}\allowbreak\; +\; 0.542\,X_{13}X_{17}\allowbreak\; +\; 1.37\,X_{13}X_{18}\allowbreak\; +\; 0.723\,X_{13}X_{19}\allowbreak\; +\; 0.321\,Y_{13}\allowbreak\; +\; 0.578\,Y_{13}Y_{14}\allowbreak\; +\; 0.542\,Y_{13}Y_{17}\allowbreak\; +\; 1.10\,Y_{13}Y_{18}\allowbreak\; +\; 0.723\,Y_{13}Y_{19}\allowbreak\; +\; 0.192\,Z_{13}\allowbreak\; +\; 0.866\,Z_{13}Z_{14}\allowbreak\; +\; 0.542\,Z_{13}Z_{17}\allowbreak\; +\; 1.65\,Z_{13}Z_{18}\allowbreak\; +\; 0.723\,Z_{13}Z_{19}\allowbreak\; +\; 0.226\,X_{14}\allowbreak\; +\; 0.463\,X_{14}X_{18}\allowbreak\; +\; 1.26\,X_{14}X_{19}\allowbreak\; +\; 0.148\,Y_{14}\allowbreak\; +\; 0.463\,Y_{14}Y_{18}\allowbreak\; +\; 1.01\,Y_{14}Y_{19}\allowbreak\; +\; 0.0797\,Z_{14}\allowbreak\; +\; 0.463\,Z_{14}Z_{18}\allowbreak\; +\; 1.52\,Z_{14}Z_{19}\allowbreak\; -\; 0.0267\,X_{15}\allowbreak\; +\; 0.862\,X_{15}X_{16}\allowbreak\; +\; 1.15\,X_{15}X_{20}\allowbreak\; +\; 0.742\,X_{15}X_{21}\allowbreak\; -\; 0.117\,Y_{15}\allowbreak\; +\; 0.690\,Y_{15}Y_{16}\allowbreak\; +\; 0.922\,Y_{15}Y_{20}\allowbreak\; +\; 0.742\,Y_{15}Y_{21}\allowbreak\; +\; 0.540\,Z_{15}\allowbreak\; +\; 1.03\,Z_{15}Z_{16}\allowbreak\; +\; 1.38\,Z_{15}Z_{20}\allowbreak\; +\; 0.742\,Z_{15}Z_{21}\allowbreak\; +\; 0.268\,X_{16}\allowbreak\; +\; 1.14\,X_{16}X_{17}\allowbreak\; +\; 0.317\,X_{16}X_{20}\allowbreak\; +\; 1.29\,X_{16}X_{21}\allowbreak\; +\; 0.336\,X_{16}X_{22}\allowbreak\; +\; 0.109\,Y_{16}\allowbreak\; +\; 0.916\,Y_{16}Y_{17}\allowbreak\; +\; 0.317\,Y_{16}Y_{20}\allowbreak\; +\; 1.03\,Y_{16}Y_{21}\allowbreak\; +\; 0.336\,Y_{16}Y_{22}\allowbreak\; +\; 0.150\,Z_{16}\allowbreak\; +\; 1.37\,Z_{16}Z_{17}\allowbreak\; +\; 0.317\,Z_{16}Z_{20}\allowbreak\; +\; 1.55\,Z_{16}Z_{21}\allowbreak\; +\; 0.336\,Z_{16}Z_{22}\allowbreak\; -\; 0.00621\,X_{17}\allowbreak\; +\; 0.892\,X_{17}X_{18}\allowbreak\; +\; 0.671\,X_{17}X_{21}\allowbreak\; +\; 0.912\,X_{17}X_{22}\allowbreak\; +\; 0.592\,X_{17}X_{23}\allowbreak\; +\; 0.288\,Y_{17}\allowbreak\; +\; 0.714\,Y_{17}Y_{18}\allowbreak\; +\; 0.671\,Y_{17}Y_{21}\allowbreak\; +\; 0.730\,Y_{17}Y_{22}\allowbreak\; +\; 0.592\,Y_{17}Y_{23}\allowbreak\; +\; 0.305\,Z_{17}\allowbreak\; +\; 1.07\,Z_{17}Z_{18}\allowbreak\; +\; 0.671\,Z_{17}Z_{21}\allowbreak\; +\; 1.09\,Z_{17}Z_{22}\allowbreak\; +\; 0.592\,Z_{17}Z_{23}\allowbreak\; -\; 0.0490\,X_{18}\allowbreak\; +\; 0.921\,X_{18}X_{19}\allowbreak\; +\; 0.625\,X_{18}X_{22}\allowbreak\; +\; 0.953\,X_{18}X_{23}\allowbreak\; +\; 0.538\,X_{18}X_{24}\allowbreak\; -\; 0.175\,Y_{18}\allowbreak\; +\; 0.737\,Y_{18}Y_{19}\allowbreak\; +\; 0.625\,Y_{18}Y_{22}\allowbreak\; +\; 0.763\,Y_{18}Y_{23}\allowbreak\; +\; 0.538\,Y_{18}Y_{24}\allowbreak\; +\; 0.405\,Z_{18}\allowbreak\; +\; 1.11\,Z_{18}Z_{19}\allowbreak\; +\; 0.625\,Z_{18}Z_{22}\allowbreak\; +\; 1.14\,Z_{18}Z_{23}\allowbreak\; +\; 0.538\,Z_{18}Z_{24}\allowbreak\; +\; 0.0666\,X_{19}\allowbreak\; +\; 0.293\,X_{19}X_{23}\allowbreak\; +\; 0.835\,X_{19}X_{24}\allowbreak\; -\; 0.00555\,Y_{19}\allowbreak\; +\; 0.293\,Y_{19}Y_{23}\allowbreak\; +\; 0.668\,Y_{19}Y_{24}\allowbreak\; +\; 0.0188\,Z_{19}\allowbreak\; +\; 0.293\,Z_{19}Z_{23}\allowbreak\; +\; 1.00\,Z_{19}Z_{24}\allowbreak\; +\; 0.105\,X_{20}\allowbreak\; +\; 0.534\,X_{20}X_{21}\allowbreak\; +\; 0.0237\,Y_{20}\allowbreak\; +\; 0.428\,Y_{20}Y_{21}\allowbreak\; +\; 0.333\,Z_{20}\allowbreak\; +\; 0.641\,Z_{20}Z_{21}\allowbreak\; +\; 0.394\,X_{21}\allowbreak\; +\; 1.10\,X_{21}X_{22}\allowbreak\; +\; 0.188\,Y_{21}\allowbreak\; +\; 0.880\,Y_{21}Y_{22}\allowbreak\; +\; 0.0831\,Z_{21}\allowbreak\; +\; 1.32\,Z_{21}Z_{22}\allowbreak\; +\; 0.140\,X_{22}\allowbreak\; +\; 1.06\,X_{22}X_{23}\allowbreak\; +\; 0.0122\,Y_{22}\allowbreak\; +\; 0.848\,Y_{22}Y_{23}\allowbreak\; +\; 0.161\,Z_{22}\allowbreak\; +\; 1.27\,Z_{22}Z_{23}\allowbreak\; +\; 0.277\,X_{23}\allowbreak\; +\; 0.669\,X_{23}X_{24}\allowbreak\; +\; 0.336\,Y_{23}\allowbreak\; +\; 0.535\,Y_{23}Y_{24}\allowbreak\; +\; 0.403\,Z_{23}\allowbreak\; +\; 0.802\,Z_{23}Z_{24}\allowbreak\; +\; 0.0637\,X_{24}\allowbreak\; -\; 0.0255\,Y_{24}\allowbreak\; +\; 0.598\,Z_{24}.\)
\par
\endgroup

\end{itemize}

\bibliographystyle{apsrev4-2}
\bibliography{references}
\end{document}